\def\bt{\begin{tabular}}          \def\et{\end{tabular}}
\def\bc{\begin{center}}                \def\ec{\end{center}}
\def\be{\begin{equation}}              \def\ee{\end{equation}}
\def\bear{\begin{eqnarray}}            \def\eear{\end{eqnarray}}
\def\nn{\nonumber}    \def\lb{\label}   \def\ci{\cite}
\def\la{\langle}      \def\ra{\rangle}     \def\l{\left}
\def\r{\right}        \def\dg{\dagger}   
\def\alf{\alpha}      \def\bet{\beta}     
     \def\sig{\sigma}    \def\tet{\theta}
\def\dlt{\delta}      \def\Dlt{\Delta}    \def\vtet{\vartheta}
\def\vphi{\varphi}    \def\gam{\gamma}    
\def\tld{\tilde}      \def\pr{\prime}
\def\ld{\ldots}       \def\ngeq{\not \geq}
                \def\rar{\rightarrow}
\def\hs{\hspace}      \def\vs{\vspace}       \def\sm{\small}

\documentstyle[12pt]{article}
\begin{document}

\begin{flushright}         Preprint INRNE-TH-97/12 (Nov 1997)\\
                                      E-print quant-ph/9711066\\
                                      (Revised April 1998)
\end{flushright} \bigskip

\bc
  {\Large \bf Barut-Girardello coherent states for $u(p,q)$ and $sp(N,R)$
             and their macroscopic superpositions }
\ec  \medskip

\centerline{ D.~A. Trifonov\footnote[0]{$^+$e-mail:
							dtrif@inrne.acad.bg}$^{+}$}
\centerline{Institute of Nuclear Research,}
\centerline{72 Tzarigradsko Chaussee,\,\,1784 Sofia, Bulgaria}
\medskip \bigskip

\bc
\begin{minipage}{14cm}
\centerline{\small {\bf Abstract}}   \bigskip

{\small
The Barut-Girardello coherent states (BG CS) representation is extended to
the noncompact algebras $u(p,q)$ and $sp(N,R)$ in (reducible) quadratic
boson realizations. The $sp(N,R)$ BG CS  take the form of multimode
ordinary Schr\"odinger cat states.  Macroscopic superpositions of
$2^{n-1}$ $sp(N,R)$ CS ($2^n$ canonical CS, $n=1,2,\ld$) are pointed
out which are overcomplete in the $N$-mode Hilbert space and the relation
between the canonical CS and the $u(p,q)$ BG-type CS representations is
established.

The sets of $u(p,q)$ and $sp(N,R)$ BG CS and their discrete superpositions
contain many states studied in quantum optics (even and odd $N$-mode CS,
pair CS) and provide an approach to quadrature squeezing, alternative to
that of intelligent states. New subsets of weakly and strongly
nonclassical states are pointed out and their statistical properties
(first- and second-order squeezing, photon number distributions) are
discussed. For specific values of the angle parameters and small amplitude
of the canonical CS components these states approaches multimode Fock
states with one, two or three bosons/photons.  It is shown that
eigenstates of a squared non-Hermitian operator $A^2$ (generalized cat
states) can exhibit squeezing of the quadratures of $A$. }

\end{minipage} \ec
\vs{1cm}

\section{Introduction}

In the recent years there has been much interest in applications and
generalizations of the Barut--Girardello (BG) coherent states (CS)
\ci{Trif94,Vourd1,Vourd2,Trif96a,Brif97a,Trif97a,Fujii}.  The
BG CS were introduced \ci{BG} as
eigenstates of the lowering Weyl operator $K_-$ of the algebra $su(1,1)$.
The BG CS representation has been used for explicit construction of
squeezed states (SS) for the generators of the group $SU(1,1)$ which
minimize the Schr\"odinger uncertainty relation for two observables
\ci{Trif94} and of eigenstates of general element of the complexified
algebra $su^C(1,1)$ \ci{Trif96a,Brif97a}.  The overcomplete families of
eigenstates of elements of a Lie algebra were called algebraic CS
\ci{Trif96a} and algebra eigenstates \ci{Brif96,Brif97a}. The idea to
construct SS for quadratures of any non-Hermitian operator $A$ as
eigenstates of complex combinations $uA+vA^\dg$ was put forward in ref.
\ci{Trif94}, where such eigenstates $|z,u,v\ra$ were  constructed for
$A=J_-$ and $A=K_-$, $J_-$ and $K_-$ being  the Weyl lowering operators of
$su(2)$ and $su(1,1)$ Lie algebras correspondingly. The $su(1,1)$ BG CS
differ from the $SU(1,1)$ group related CS (see \ci{KlSk} and references
therein):  the highest weight vector is the only one common state, while
the $su(1,1)$ algebra related CS $|z,u,v;k\ra$ of ref. \ci{Trif94} contain
the whole set of $SU(1,1)$ group related CS with symmetry. The general set
of algebra related CS always contains  the corresponding group related CS
with symmetry as a subset.

Passing to other algebras it is initially important to construct the
eigenstates of Weyl lowering operators, which is a direct extension of the
BG definition of $su(1,1)$ CS to the desired algebra.  The aim of the
present work is to construct BG type CS for the symplectic algebra
$sp(N,R)$ and its subalgebras $u(p,q),\,\,p+q=N$,  in the quadratic boson
representation. This  $sp(N,R)$ representation is of importance
in various field of physics \ci{BaRo,Mosh,Tod}.  Here $N$ is the dimension
of Cartan subalgebra, while the dimension of $sp(N,R)$ is $N(2N+1)/2$,
$N=1,2,\ld$, \ci{BaRo}.

We establish that the $sp(N,R)$ BG CS in quadratic boson
representation take the form of superpositions of two multimode canonical
CS \ci{KlSk} $|\vec{\alf}\ra$ and $|\!-\!\vec{\alf}\ra$ (equation
(\ref{spNRbgcssolu})). A subset of these states is found which is
overcomplete in the whole Hilbert space ${\cal H}$ of the $N$-mode system.
Recall that the corresponding $Sp(N,R)$ group related CS are not
overcomplete in ${\cal H}$ since the representation is reducible.  This
property is a particular case of a quite general result
of the overcompleteness of eigenstates of powers $A_j^{2^n}$ of
non-Hermitian $A_j$, $j=1,2,\ld$, provided the eigenstates $|\vec{z}\ra$,
$\vec{z} = (z_1,\ld,z_N)$, of all $A_j$ are overcomplete with respect to a
measure independent of the phases of $z_j$ (section 3 and appendix A.2).

Macroscopic superpositions of two canonical CS are called (ordinary)
Schr\"odinger cat states \ci{cats1,cats2}. The set of the $sp(N,R)$ BG CS
includes several subsets of ordinary cat states, which are extensively
studied in quantum optics (see \ci{cats1,cats2} and references therein).
We introduce {\it multimode squared amplitude Schr\"odinger cat states} as
macroscopic superpositions of two $sp(N,R)$ BG CS.  Unlike the
ordinary cat states these superpositions, which eventually become
combinations of four $N$ mode canonical CS, can exhibit amplitude and squared
amplitude quadrature squeezing (first- and second-order quadrature
squeezing or linear and quadratic squeezing) \ci{Hill}, and other
nonclassical properties. Families of weakly and strongly nonclassical
\ci{Arv1} cat states  are pointed out as macroscopic superpositions of two
$Sp(N,R)$ CS.  There are states  in these families
that tend to multimode Fock states with $0,1,2$ or $3$ photons as the
amplitude of their canonical CS components approaches zero.  We note that,
unlike the case of Robertson (Schr\"odinger) intelligent states
\ci{Trif94,Trif97a}, the cat state squeezing can not be arbitrarily
strong.

Recently \ci{Fujii} the BG type CS have been constructed for the $u(N-1,1)$
algebra.  Here we construct overcomplete families of states for $u(p,q)$,
$p+q = N$, and the related resolution unity measures as well. We show that
the  `pair CS' \ci{Agar} are in fact the $u(1,1)$ BG CS $|z;k\ra$ for
$k=1/2,1\ld$, while the `two-mode Schr\"odinger cat states' of
\ci{Ger2} are particular case of our $u(p,q)$ multimode squared amplitude
cat states (\ref{u(p,q)cats}).

The paper is organized as follows.
In section 2 a concise review of the properties of BG CS representation
and its relations to the canonical one- and two-mode CS representation (or
Fock-Bargman representation)\ci{KlSk} is given. An explicit relation
between the two-mode canonical CS and the BG CS representations is
obtained.  Using this relation on can more easily establish the
coincidence between the generalized intelligent states (IS) $|z,u,v;k\ra$
\ci{Trif94} and many other one and two-mode states, constructed by other
authors as eigenstates of $ua^2 + va^{\dg 2}$  or $uab + va^\dg b^\dg$
\ci{Prak,Ger1,Shanta}. For example, for real $u,\,v$ the states
$|z,u,v;k\!=\!1/2\ra$ coincide with the `pair excitation-deexcitation CS'
\ci{Prak}, while for real $u,\,v$ and $k=(1+|q|)/2$ they are identical to
the `two-mode intelligent $SU(1,1)$ CS' \ci{Ger1}.

In section 3 the BG CS are extended explicitly to the algebra $sp(N,R)$ in
the (reducible) quadratic boson representation and
overcomplete in whole ${\cal H}$ families of such states are constructed.
Overcomplete families of eigenstates of the power $2n$ of Weyl operators
$a_ia_j$ are also built up. These states take the form of macroscopic
superpositions with $2^n$ canonical CS components.  The $u(p,q)$ BG type CS
are considered in section 4 (overcompleteness, resolution unity measure,
particular cases and relation of their analytic representation to that of
canonical CS).  In section 5 the statistical properties (weak and strong
nonclassicality, amplitude and squared amplitude quadrature squeezing,
sub- and super-Poissonian photon statistics) of the constructed $sp(N,R)$
algebra related CS and their superpositions are discussed and illustrated
by several graphics. Our analysis shows that photon
number oscillations are not necessary characteristics of nonclassicality
of quantum states (neither are they sufficient \ci{Arv2}).  We note the
main difference between squeezing in intelligent SS \ci{Trif94,Trif97a,NT}
and in cat-type SS and construct a second kind multimode squeeze operator
as a map from CS $|\vec{\alf}\ra$ to a set of cat-type multimode SS.  In
the appendix several statements of the main text are proved.
\vspace{5mm}

\section{The Barut-Girardello coherent states}

The property of canonical CS $|\alf\ra$ \ci{KlSk} to be
eigenstates of photon number lowering operator $a$, $a|\alf\ra =
\alf|\alf\ra$ ($\alf$ is complex number, $[a,a^\dg] = 1$) was extended by
Barut and Girardello \ci{BG} to the case of Weyl lowering operator $K_-$
of $su(1,1)$ algebra.  Here we briefly review some of their properties.
The defining equation is

\be\lb{bgcsdef}  
K_-|z;k\ra = z|z;k\ra,
\ee
where $z$ is (complex) eigenvalue and $k$ is Bargman index. Here,
and in \ci{Trif94}, we introduced $k=-\Phi$ as a second label of the state
and replaced the BG $z$ with $z/\sqrt{2}$. For discrete series
$D^{(\pm)}(k)$ the parameter $k$ takes the values $\pm1/2,\,\pm1,\,\ld$.
The Cartan--Weyl basis operators $K_{\pm}=K_1\pm iK_2,\,\, K_3$ of
$su(1,1)$ obey the relations

\be\lb{su11comrel}   
  [K_3,K_\pm] = \pm K_\pm, \,\,\,\, [K_-,K_+] = 2K_3,
\ee
with the Casimir operator $C_2 = K_3{}^2 -(1/2)[K_-K_+ + K_+K_-] =
k(k-1)$.  The expansion of these states over the orthonormal basis of
eigenstates $|k+n,k\ra$ of $K_3$ ($K_3|n+k,k\ra = (n+k)|n+k,k\ra$, $n =
0,1,2,\ld$) is

\begin{eqnarray}\lb{bgcsexpan}   
|z;k\rangle =  {\cal N}_{BG}(|z|,k)
\sum_{n = 0}^{\infty}
\frac{z^{n}}{\sqrt{n!\Gamma(2k+n)}} |n+k,k\rangle \equiv
{\cal N}_{BG}(|z|,k)|\!|z;k\ra, \\
{\cal N}_{BG}(|z|,k) =  [\Gamma(2k)/{}_0F_1(2k;|z|^2)]^{\frac{1}{2}} =
\frac{|z|^{k-1/2}}{\sqrt{I_{2k-1}(2|z|)}},\nonumber
\end{eqnarray}
where $_0F_1(c;z)$ is the confluent hypergeometric function, $I_\nu(z)$ is
the modified Bessel function of the first kind, and $\Gamma(z)$ is the
gamma function \ci{Stegun}.  The above BG states $|z;k\ra$ are normalized
to unity.  Their scalar product is

\be\lb{bgcsscalprod}  
\la k;z|z^\pr;k \ra =
{}_0F_1(2k;z^*z^\pr)\l[_0F_1(2k;|z|^2)\,_0F_1(2k;|z^\pr|^2)\r]^{-\frac{1}{2}},
\ee
and they resolve the unity (the identity operator),

\be\lb{bgcsresolu}   
\int d\mu(z,k) |\!|z;k\ra\la k;z|\!| = 1_k, \qquad d\mu(z,k) =
  \frac{2}{\pi}|z|^{2k-1} K_{2k-1}(2|z|)\, d^{2}\! z ,
  \ee
where $K_{\nu}(x)$ is the modified Bessel function of the third kind. Note
that $|\!|z;k\ra = {\cal N}_{BG}^{-1}|z;k\ra$, while  in \ci{BG} these
nonnormalized CS were denoted as $\Gamma(2k)^{-1/2}|z\ra$ (note also the
misprint in \ci{BG}:  in formula for the measure function
$\sig(r)$ one should replace $K_{\Phi+\frac 12}(2\sqrt{2}r)$
by $K_{2\Phi+1}(2\sqrt{2}r)$ \ci{Vourd1}).  Owing to the above
overcompleteness property any state $|\Psi\ra$ can be correctly
represented by the analytic function

\be\lb{F_BG}  
F_{BG}(z,k;\Psi) = \la k,z^{\ast}|\Psi\ra/{\cal N}_{BG}(|z|,k) = \la
k,z^\ast|\!|\Psi\ra,
\ee
which is of the growth $(1,1)$.  The orthonormalized states
$|k+n,k\ra$ are represented by monomials $z^n/\sqrt{n!\Gamma(2k+n)}$ (we
note a misprint in these monomials in \ci{Trif96a} and \ci{Trif97a}: $k$
should be replaced by $2k$).  The operators ${K}_{\pm}$ and ${K}_{3}$ act
in the space ${\cal H}_k$ of analytic functions $F_{BG}(z,k)$ as linear
differential operators

\be\lb{bgKrep}  
 K_{+} =  z,  \quad K_{-} = 2k \frac{d}{dz} + z \frac{d^{2}}{dz^{2}},
 \quad K_{3} = k + z \frac{d}{dz}  .
\ee
This analytic representation has been used to explicitly construct
eigenstates $|z,u,v;k\ra$ of complex combinations $uK_- + vK_+$ in paper
\ci{Trif94}.

Barut and Girardello have established their continuous representation for
the discrete series $D^{\pm}(k)$, $k=\pm1/2,\pm1,\ld$. However, by
inspection of their construction one can  easily see that it also holds
for {\it reducible representations} and for $1/2>|k| > 0$  - one
only has to keep in mind that the quantity $1_k$ in the overcompleteness
relation (\ref{bgcsresolu}) is the identity operator in the subspace
${\cal H}_k$, {\it where $su(1,1)$ acts irreducibly}.  The proof consists
of two observations (for concreteness we take $D^+(k)$):  (a) The
expansions (\ref{bgcsexpan}) are convergent and represent normalized
states for $k \geq 0$, provided $|k,k+n\ra$ are orthonormalized; (b) The
BG measure $d\mu(z,k)$ resolves the unity operators by means of $|z;k\ra$
for $k \geq 0$ provided the orthonormalized set of $|k,k+n\ra$ is
complete.

It is well known that the $su(1,1)$ algebra has one- and two-mode
quadratic boson representations, which are reducible in the spaces of
states of one- and two-mode systems correspondingly. The one-mode
realization of $su(1,1)$ is

\be\lb{K_i N=1}  
K_- = \frac{1}{2}a^2,\,\, K_2 =
\frac{1}{2}a^{\dg 2},\,\, K_3 = \frac{1}{4}(2a^\dg a +1).
\ee
Its quadratic Casimir operator $C_2$ equals $-3/16$, $C_2 = K_3^2 - K_1^2
- K_2^2 = k(k-1)$, the Bargman index being $k=1/4,\,3/4$. The two-mode
representation

\be\lb{K_i N=2}  
K_-  = a_1a_2,\,\, K_+ =  a^\dg_1a^\dg_2,\,\,K_3 = \frac 12(a^\dg_1a_1 +
a^\dg_2a_2 + 1).
\ee
is highly reducible (completely reducible), its irreducible components
being just the representations from the  discrete series $D^+(k)$,
$k=1/2,1,\ld $.  The whole space ${\cal H}$ of the two-mode system states
is a direct sum of the irreducible modules ${\cal H}_k$. In these
realizations the operators $uK_- + vK_+$, which were diagonalized in ref.
\ci{Trif94}, read $ua^2+va^{\dg 2}$ and $ua_1a_2 + va^\dg_1a^\dg_2$.

The Heisenberg--Weyl algebras $h_1$ and $h_2$, spanned by
$1,\,a_1,\,a^\dg_1$ and $1,\,a_1,\,a^\dg_1,\,\,a_2,\,a^\dg_2$
correspondingly, act irreducibly in the state spaces of one- and two-mode
systems. The related families of CS $|\alf\ra$ and $|\alf_1,\alf_2\ra$ are
overcomplete and realize the continuous representations, which proved to
be very efficient \ci{KlSk}.  Therefore it is important to establish the
relation between BG CS and the canonical CS representations.  In the
canonical CS representation every state $|\Psi\ra$ is represented by an
entire analytic function $F_{CCS}(\alf_1,\alf_2;\Psi)$ of growth (1/2,2),

\be\lb{F_CCS 1}   
F_{CCS}(\alf_1,\alf_2;\Psi) =
\exp\l(\mbox{\sm $\frac 12$}(|\alf_1|^2+|\alf_2|^2)\r)\la
\alf^*_1,\alf^*_2|\Psi\ra.
\ee
In the one-mode case $F_{CCS}(\alf) = \exp(\mbox{\sm $\frac
12$}|\alf|^2)\la \alf^*|\Psi\ra$. The eigenvalue properties of the BG CS
and canonical CS and the realizations (\ref{K_i N=1}) and (\ref{K_i N=2})
suggest that the canonical CS representation of a state $|\Psi\ra \in
 {\cal H}_k$ should be obtained (up to a common factor) from its BG
representation by means of substitution $z=\alf^2/2$ for the one-mode
system and $z=\alf_1\alf_2$ for the two-mode system, and this is the case.
The corresponding relation between the two representations of the one-mode
system states was written down in \ci{Vourd1},

\be\lb{CCS-BG N=1}  
F_{CCS}(\alf;\Psi) = \pi^{\frac 14}\l[F_{BG}(\mbox{\sm $\frac
12$}\alf^2,k\!=
\!\mbox{\sm $\frac 14$}) + \frac{1}{\sqrt{2}}\alf F_{BG}(\mbox{\sm $\frac
12$}\alf^2,k\!=\!\mbox{\sm $\frac 34$})\r].
\ee
If $|\Psi\ra$ is even (odd) state, then the second (the first) term is
vanishing.
{}For the two-mode system states the relation between
$F_{CCS}$ and $F_{BG}$, defined above, is found in the form (proof in the
appendix A.1)

\be\lb{CCS-BG N=2}   
F_{CCS}(\alf_1,\alf_2;\Psi) = F_{BG}(z,k\!=\!\mbox{\sm $\frac 12$};\Psi) +
\sum_{k\geq 1}
\l(\alf_1^{2k-1}+\alf_2^{2k-1}\r)F_{BG}(z,k;\Psi),\, z\!=\!\alf_1\alf_2.
\ee
Using these relations one can establish the coincidence between states,
obtained in BG analytic representations and other familiar states. For
example, the known one-mode even/odd CS $|\alf\ra_\pm$ coincide with the
BG CS $|z;1/4\ra$ and $|z;3/4\ra$ \ci{Vourd1,Basu}, while the generalized
IS $|z,u,v;k\ra$,  constructed in \ci{Trif94} using BG representation, for
$k=1/4,\,3/4$ are the same as the eigenstates of $ua^2 + va^{\dg 2}$,
constructed for real $u,\,v$ in \ci{NT} and for complex $u,\,v$ in
\ci{Trif95,Trif96a,Trif96b,Brif96} using the canonical CS representation. For
$k\geq1/2$ the $|z,u,v;k\ra$ with real $u,\,v$ can be identified with
two-mode $SU(1,1)$ states of \ci{Agar,Prak,Ger1}. All $SU(1,1)$
states of \ci{Puri,Perina} can be found in the general family of
$su^C(1,1)$ algebra related CS  $|z,u,v,w;k\ra$ constructed in
\ci{Trif96a,Trif97a,Brif97a}.

In conclusion to this section it is worth noting that the $SU(1,1)$ group
related CS \ci{KlSk} provide another analytic (in the unit disk) \ci{Ben}
representation of Hilbert space which has been shown \ci{Vourd1} to be
related to the BG representation through a Laplace transform. It is also
worth making a note about the notations: the BG CS are eigenstates of
lowering operator $K_- = K_1 -i K_2$, which belongs to the complexified
algebra $su^C(1,1)$. Therefore we could denote such states as $su^C(1,1)$
algebra related CS.   However, usually  when one deals with such simple
complex combination as Weyl lowering/raising operators of an algebra $L$
($K_\pm$ for $su(1,1)$) one writes $L$ instead of $L^C$ ($su(1,1)$ instead
of $su^C(1,1)$). For brevity we follow here this convention for BG CS for
Lie algebras.  Continuous families of eigenstates of general element  of
$su^C(1,1)$ have been considered and called $su^C(1,1)$ algebraic CS
\ci{Trif96a} or $SU(1,1)$ algebra eigenstates \ci{Brif97a}.
An other motivation of the new term `algebra related CS' is the following
property of the BG CS $|z;k\ra$: unlike the $h_n^C$ algebra CS this family
can not be represented in the form of group related CS either for the
group $SU(1,1)$ or for the group of automorphysm Aut$(su^C(1,1)) \ni
SU(1,1)$ \ci{Trif97b}.
\bigskip

\section{ The BG CS for $sp(N,R)$ }

The BG CS for semisimple Lie algebras can be naturally defined as eigenstates
of mutually commuting Weyl lowering (or raising) operators $E_{\alf^\pr}$
($E^\dg_{\alf^\pr}$) \ci{BaRo}):

\be\lb{gen-bgcs}   
E_{\alf^\pr}|\vec{z}\ra = z_{\alf^\pr}|\vec{z}\ra.
\ee

This definition can be extended to any algebra, where lowering/raising
operators exist.  We shall consider here the simple Lie algebra $sp(N,R)$
(the symplectic algebra of rank $N$ and dimension $N(2N+1)$). We redenote
the Cartan-Weyl basis as $E_{ij}, E^\dg_{ij}, H_{ij}$ ($i,j =
1,2,\ld,N$, $E_{ij}=E_{ji}$, $H_{ij}^\dg = H_{ji}$), and write the
$sp(N,R)$ commutation relations

\be\lb{spNRcomrel}  
\bt{l}
$[E_{ij},E_{kl}] = [E^\dg_{ij},E^\dg_{kl}] = 0$,\\
$[E_{ij},E^\dg_{kl}] = \delta_{jk}H_{il} + \delta_{il}H_{jk} +
\delta_{ik}H_{jl} + \delta_{jl}H_{ik}$,\\
$[E_{ij},H_{kl}] = \delta_{il}E_{jk} + \delta_{jl}E_{ik}$,\\
$[E^\dg_{ij},H_{kl}] = -\delta_{ik}E^\dg_{jl} - \delta_{jk}E^\dg_{il}$,\\
$[H_{ij},H_{kl}] = \delta_{il}H_{kj} - \delta_{jk}H_{il}$.  \et
\ee \vs{3mm}

The BG CS $|\{z_{kl}\}\ra$ for $sp(N,R)$ are defined as eigenstates of
$E_{ij}$,

\be\lb{spNRbgcsdef1}  
E_{ij}|\{z_{kl}\}\ra = z_{ij}|\{z_{kl}\}\ra,\quad i,j=1,2,\ld,N.
\ee
Let us note that the Cartan subalgebra is spanned by $H_{ii}$ only and
$H_{i,j\neq i}$ are also Weyl lowering and raising operators as all
$E_{ij}$ are:  we have simply separated the {\it mutually commuting lowering
operators $E_{ij}$}.  We shall construct explicitly the $sp(N,R)$ BG CS
for the quadratic boson rep, which is realized by means of
the operators

\be\lb{spNRbosrep}  
E_{ij} = a_ia_j,\quad E^\dg_{ij} = a^\dg_ia^\dg_j,\quad
H_{ij} = \frac{1}{2}(a^\dg_ja_i + a_ia^\dg_j),
\ee
where $a_i,\,a^\dg_i$ are $N$ pairs of boson annihilation and creation
operators. These operators act irreducibly in the subspaces ${\cal
H}^{\pm}$ spanned by the number states $|n_1,\ld,n_N\ra$ with
even/odd $n_{tot}\equiv n_1+n_2+\ld+n_N$. The whole space
${\cal H}$ of the $N$ mode system is a direct sum of ${\cal H}^{\pm}$.

The $sp(N,C)$ is the complexification of $sp(N,R)$ and therefore the
Hermitian quadratures of the above operators span over $C$ the
$sp(N,C)$ algebra.  In the case of $N=1$ one obtains from
(\ref{spNRbosrep}) the three operators $K_{\pm,3}$ which close
$sp(1,R)\sim su(1,1)$ (see equation (\ref{K_i N=1})).
We see that eigenstates of $a^2$ (the known even/odd states $|\alf\ra_\pm$
in quantum optics \ci{cats1}) are $sp(1,R)$ BG CS for $k=1/4,3/4$.

One general property of $sp(N,R)$ CS $|\{z_{kl}\}\ra$ for the
representation (\ref{spNRbosrep}) is that they depend effectively on $N$
complex parameters $\alf_j$ (not of $N^2+N$ as one might expect).  Indeed,
using the boson commutation relations $[a_i,a_j] = 0$ and the
definition (\ref{spNRbgcsdef1}) we easily derive

\be\lb{zijfactoriz}  
z_{ij}z_{kl} = z_{ik}z_{jl} = z_{il}z_{jk},
\end{equation}
wherefrom we find the factorization of the eigenvalues $z_{ij}$,

\be\lb{factoriz} 
z_{ij} = \alf_i\alf_j, \quad \alf_i,\,\alf_j \in C. \ee

Therefore in the above boson representation the definition
(\ref{spNRbgcsdef1}) is rewritten as

\be\lb{spNRbgcsdef2}  
a_ia_j|\{\alf_k\alf_l\}\ra = \alf_i\alf_j|\{\alf_k\alf_l\}\ra,
\quad i,j=1,2,\ld,N.
\ee

The general solution to this system of equations is most easily obtained
in the canonical CS representation \ci{KlSk}. In Dirac notations the solution
reads

\be\lb{spNRbgcssolu}  
|\{\alf_k\alf_l\};C_+,C_-\ra = C_+(\vec{\alf})|\vec{\alf}\ra +
C_-(\vec{\alf})|-\vec{\alf}\ra \equiv |\vec{\alf};C_+,C_-\ra,
\ee
where $|\vec{\alf}\ra$ are multimode canonical CS, $\vec{\alf} =
(\alf_1,\alf_2,\ld,\alf_N)$ and $C_\pm(\vec{\alf})$ are
arbitrary functions, subjected to the normalization condition
($|\vec{\alf}|^2 = \vec{\alf}\cdot\vec{\alf}= |\alf_1|^2 + \ld
+|\alf_N|^2$)

\be\lb{normcond1}  
|C_+(\vec{\alf})|^2 + |C_-(\vec{\alf})|^2 +
2{\rm Re}(C_-C^\dg_+)\,N(|\vec{\alf}|) = 1, \quad
N(|\vec{\alf}|)=\la\pm\vec{\alf}|\mp\vec{\alf}\ra = e^{-2|\vec{\alf}|^2}.
\end{equation}
Thus the families of states $|\vec{\alf};C_+,C_-\ra$ represent {\it the
whole set of $sp(N,R)$ BG CS} for the representation (\ref{spNRbosrep}).
They have the form of macroscopic superpositions of multimode canonical CS.
Macroscopic superposition of two canonical CS are also called
Schr\"odinger cat states \ci{cats1,cats2}, which we shall refer to as
ordinary Schr\"odinger cat states. The set of (\ref{spNRbgcssolu}) is the
most general family of superpositions of multimode CS $|\vec{\alf}\ra$ and
$|-\vec{\alf}\ra$.

The large family of $sp(N,R)$ CS (\ref{spNRbgcssolu}) contains many
known, in quantum optics, subsets of states \ci{cats1,cats2} and
many other not yet studied. Let us  point out some of the well known
particular subsets of (\ref{spNRbgcssolu}). The limiting cases of
$C_-=0$ or $C_+=0$ recover the overcomplete family of multimode canonical CS, and
and $C_-=\pm C_+$ produces the ordinary multimode even/odd CS \ci{cats2}.

For the one-mode system ($N=1$) several cases  of the superpositions of
two canonical CS (\ref{spNRbgcssolu}) are thoroughly studied (for example,
see \ci{cats1} for $N=1$ and \ci{cats2} for any  $N$).  Nevertheless, as
far as we know, even in the one dimensional case no family of
Schr\"odinger cat states was pointed out which is overcomplete in the
strong sense in whole ${\cal H}$.  Here we provide such families for any
$N$.

Consider in (\ref{spNRbgcssolu}) the choice of

\be\lb{Cpm,phi}   
C_+ = \cos\varphi,\,\, C_- = i \sin\varphi,
\ee
which clearly satisfy the normcondition (\ref{normcond1}) for any
angle $\varphi$,

\be\lb{alf;vphi1}   
|\vec{\alf};\vphi\ra = \cos\vphi|\vec{\alf}\ra + i\sin\vphi |-\vec{\alf}\ra.
\ee
In Fock basis (number states $|n_1,\ld,n_N\ra$) we have the expansion

\be\lb{alf;vphi2}   
|\vec{\alf};\varphi\ra = e^{-|\vec{\alf}|^2/2} \sum_{n_i=0}^{\infty}
\frac{\alf_1^{n_1}\ld\alf_N^{n_N}e^{i\varphi
(-1)^{n_1+\ld+n_N}}}{\sqrt{n_1!\ld n_N!}}|n_1,\ld,n_N\ra.
\ee
Using direct calculations we find that these states resolve the unity
operator for any $\varphi$ and thereby provide an analytic representation
in the whole ${\cal H}$,

\be\lb{alf;vphi resolu}  
1 = \frac{1}{\pi^N}\int d^2\vec{\alf}
| \vec{\alf};\varphi\ra \la\varphi; \vec{\alf}|, \quad
d^2\vec{\alf} = d{\rm Re}\alf_1\,d{\rm Im}\alf_1 \ld d{\rm Re}
\alf_N\,d{\rm Im}\alf_N .
\ee
States $|\Psi\ra$ are represented by  functions

$$f_\Psi(\vec{\alf},\vphi) = e^{|\vec{\alf}|^2/2} \la
\vphi,\vec{\alf}^{\ast}|\Psi\ra, $$
on which the operators $a_j$ and $a^\dg_j$ act as

\be\lb{aa^dagrep}   
a_j = P_\vphi \alpha_j, \quad a^\dg_j = P_\vphi\frac{\partial}
{\partial \alpha_j},
\ee
where $P_\vphi$ acts as inversion operator with
respect to $\vphi$:  $P_\vphi f(\vphi) = f(-\vphi)$. At $\vphi = 0, \pi$
the multimode canonical CS representation $a_j=\alf_j,\, a^\dg_j =
\partial/\partial\alf_j$ is recovered.

The notations of (\ref{spNRbgcsdef1}) enable us to construct eigenstates
of squared Weyl operators $E^2_{ij}$ (in any representation) as macroscopic
superpositions of $sp(N,R)$ BG CS in the form ($z_{ij}$ are eigenvalues of
$E_{ij}$)

\be\lb{spNRcats} 
|\{z_{kl}\}; D_+,D_-\ra = D_+(\{z_{ij}\})|\{z_{kl}\}\ra +
D_-(\{z_{ij}\})|\{-z_{kl}\}\ra,
\ee
where  the functions $D_{\pm}(\{z_{ij}\})$ have to be subjected to the
normalization condition  (supposing $\la\{z_{kl}\}|\{z_{kl}\}\ra = 1$)

\be\lb{normcond2}  
|D_+|^2 + |D_-|^2 + D_-D^*_+\la\{z_{kl}\}|\{-z_{kl}\}\ra +
D^*_-D_+\la\{-z_{kl}\}|\{z_{kl}\}\ra = 1.
\ee
In the quadratic boson representation (\ref{spNRbosrep}) these states take
the form

\bear\lb{SA cats}   
|\{\alf_k\alf_l\};D_+,D_-\ra  = D_+|\{\alf_i\alf_j\};C_+,C_-\ra +
D_-|\{-\alf_i\alf_j\};C_+,C_-\ra \nn \\
\equiv |\vec{\alf};C_+,C_-, D_+,D_-\ra
\eear
and can be termed {\it multimode squared amplitude Schr\"odinger cat
states}. They are expected to exhibit linear and quadratic
squeezing and other nonclassical properties. In view of
(\ref{spNRbgcssolu}) the states (\ref{SA cats}) are eventually expressed
in terms of superpositions of four multimode canonical CS.

In conclusion to this section we note that the overcomplete
family of states $|\vec{\alf};\vphi\ra$ admits $n$ angles generalization:
by means of $n$ angles $\vphi_k$, $k = 1,2,\ld,n$, $n$ being positive
integer, one can construct macroscopic superpositions of $2^n$ CS
$|\vec{\alf}\ra$ (or, equivalently, superpositions of $2^{n-1}$ $sp(N,R$ CS
of the type $|\vec{\alf};\vphi\ra$),   which are overcomplete and resolve
the unity with respect to the same measure $\pi^{-N} d^2\vec{\alf}$,

\be\lb{2nCCS sup}   
|\vec{\alf};\vphi_1,\ld,\vphi_n\ra = \cos\vphi_n
|\vec{\alf};\vphi_1,\ld,\vphi_{n-1}\ra +
i\sin\vphi_n|-\vec{\alf};\vphi_1,\ld,\vphi_{n-1}\ra,
\ee
\be\lb{alf;Nvphi resolu} 
1 = \frac{1}{\pi^N}\int d^2\vec{\alf}|\vec{\alf};\vphi_1,\ld,\vphi_n\ra
\la \vphi_n,\ld,\vphi_1;\vec{\alf}|.
\ee
In  every component state in $|\vec{\alf};\vphi_1,\ld,\vphi_n\ra$ the
parameters $\alf_i$ are on a circle with radius $|\alf_i|$.  {}For $n=0$
we have CS $|\vec{\alf}\ra$, for $n=1$ the states (\ref{alf;vphi1}) are
reproduced.
\,\,$|\vec{\alf};\vphi_1,\ld,\vphi_n\ra$ are easily seen to be
eigenvectors of $(a_ia_j)^{2^{n-1}}$, and not of $(a_ia_j)^m$,
$m<2^{n-1}$, unless $\vphi_k$ are integer multiples of $\pi/2$. In the
one mode case ($N=1$) $|\alf;\vphi_1,\ld,\vphi_n\ra$ are eigenstates of
$a^{2^n}$.  Eigenstates of $a^{2k}$ for $k=1,2,\ld$, can be easily
constructed as superpositions of $sp(1,R)$ CS. Here we proved their
overcompleteness for $2k = 2^n = 2,4,8,32\ld$. Some eigenstates of
powers of $a^k$, $k>2$, have been considered in refs. \ci{Ger3}.
Multicomponent macroscopic superpositions of canonical CS (one mode only
so far) are intensively studied in quantum optics (with the final aim
being the production of Fock states) \ci{Vacc,Bose,Jans}.

The above result (31) is a particular case of a general theorem, proved
in appendix A.2, concerning the overcompleteness of common eigenstates
of powers of $N$ non-Hermitian operators $A_j^{2^n}$, $j=1,\ld,N$,
$n=1,2,\ld$, and valid for the case of $N$ mode  canonical CS and $sp(N,R)$ BG CS
as well.
\bigskip

\section{BG CS for the algebra $u(p,q)$}

The algebras $u(p,q)$, $p+q=N$, are real forms of $sl(N,C)$ and they are
subalgebras of $sp(N,R)$ \ci{BaRo}.  Therefore the BG CS for $u(p,q)$
should be obtained from $sp(N,R)$ CS by a suitable restrictions. In this
section we consider these problems in greater detail in the boson
representation (\ref{spNRbosrep}).

The following subset of operators of (\ref{spNRbosrep}) close
the $u(p,q)$ algebra (or $u^C(p,q)$ if one consider non-Hermitian
linear combinations of the operators below) \ci{BaRo},

\be\lb{u(p,q)bosrep}    
E_{\alf\mu} = a_\alf a_\mu,\quad E^\dg_{\alf\mu} = a^\dg_\alf a^\dg_\mu,\quad
H_{\alf\beta} = \frac{1}{2}(a^\dg_\alf a_\beta + a_\beta a^\dg_\alf), \quad
H_{\mu\nu} = \frac{1}{2}(a^\dg_\mu a_\nu + a_\nu a^\dg_\mu),
\ee
where we adopted the notations $\alf,\,\beta,\,\gamma  = 1,\ld,p$,
$\,\,\, \mu,\,\nu = p+1,\ld,p+q,\,\,\,\,p+q=N$ (while $i,j,k,l =
1,2,\ld,N$).  For $p= 1 =q$ the three standard $su(1,1)$ operators
$K_\pm,\,K_3$ are
$K_- = E_{12} = a_1a_2,\,\, K_+ =
E^\dg_{12}=a^\dg_1a^\dg_2,\,\,K_3 = (a^\dg_1a_1 +
a^\dg_2a_2 + 1)/2$.
The subsets of Hermitian operators
\be\lb{u(p),u(q)}   
\bt{l}
$M^{(p)}_{\alf\bet} =\frac{1}{2}(H_{\alf\bet}+H_{\bet\alf}-\dlt_{\alf\bet}),
\,\, \tld{M}^{(p)}_{\alf\bet} = i(H_{\bet\alf}- H_{\alf\bet})$,\\
$M^{(q)}_{\mu\nu} =\frac{1}{2}(H_{\mu\nu}+H_{\nu\mu}-\dlt_{\mu\nu}),
\,\, \tld{M}^{(q)}_{\mu\nu} = i(H_{\nu\mu}- H_{\mu\nu})$ \et
\ee
realize representations of compact subalgebras $u(p)$ and $u(q)$
correspondingly.  The $u(p,q)$   algebra (\ref{u(p,q)bosrep}) acts
irreducibly in the subspaces of eigenstates of the Hermitian operator $L$,

\be\lb{L}    
L = \sum_\alf M^{(p)}_{\alf\alf} -
\sum_\mu M^{(q)}_{\mu\mu} = \sum _\alf H_{\alf\alf} - \sum_\mu
H_{\mu\mu} - (p-q)/2.  \ee

This is the linear-in-generators Casimir operator and the higher Casimirs
here are expressed in terms of $L$ \ci{Tod}).  Denoting the eigenvalue of
$L$ by $l$ we have the expansion ${\cal H} = \sum_{l=-\infty}^{\infty}
\oplus{\cal H}_l$. The representations corresponding to $\pm l$ are
equivalent (but the subspaces ${\cal H}_{\pm l}$ are orthogonal). We note
that $L = \sum_\alf a^\dg_\alf a_\alf - \sum_\mu a^\dg_\mu a_\mu$, and $l
= 0,\pm1,\ld$.

The commuting Weyl lowering operators of $u(p,q)$ are $E_{\mu\gamma} =
a_\mu a_\gam$, $\gam = 1,2,\ld,p$, $\mu=p+1,p+2,\ld,p+q=N$. We have
proved in the above that eigenvalues of the product of two boson
destruction operators are factorized.  Therefore the $u(p,q)$ BG CS in the
above boson representation can be defined as
$|\{\alf_\bet\alf_{\nu}\};l,p,q\ra$,

\be\lb{u(p,q)bgcsdef}    
\bt{l}$a_\mu a_\gam |\{\alf_\bet\alf_\nu\};l,p,q\ra = \alpha_\mu
\alpha_\gam |\{\alf_\bet\alf_\nu\};l,p,q\ra,$\\
										\\[-2mm]
$\gam=1,\ld,p,\,\,\,\,\mu=p+1,\ld,p+q,$ \et
\ee
where $\alf_\mu$ and $\alf_\gam$ are arbitrary complex numbers.
 We put $|\!|\vec{\alf};l,p,q\ra =
|\!|\{\alf_\bet\alf_\nu\};l,p,q\ra$, denoting by $|\!|\Psi\ra$ a
nonnormalized (but normalizable) state, while $|\Psi\ra$ is normalized to
unity. Solutions to the above equations can be written in the form

\be\lb{u(p,q)bgcssolu}   
|\!|\vec{\alf};l,p,q\ra =
\sum_{\tld{n}_p-\tld{n}_q=l}
\frac{\alf_1^{n_1}\ld\alf_{N-1}^{n_{N-1}}\alf_N^{\tld{n}_p-\tld{n}^\pr_q-l}}
{\sqrt{n_1! \ld n_{N-1}!(\tld{n}_p-\tld{n}^\pr_q-l)!}}|n_1,...,n_{N-1};
\tld{n}_p - \tld{n}^\pr_q-l\ra,
\ee
where $\alpha_i$, $i=1,\ld,N$, are arbitrary complex parameters,
$\tld{n}_p = \sum_\alf n_{\alf}$, $\tld{n}_q = \sum_\mu n_\mu$,
$\tld{n}^\pr_q = \tld{n}_q - n_N$ and $l = \tld{n}_p -
\tld{n}_q$. In (\ref{u(p,q)bgcssolu}) summation is over all $n_i =
0,1,2,\ld$ provided $\tld{n}_p-\tld{n}_q=l={\rm const}$.

If we multiply $|\!|\vec{\alf};l,p,q\ra$ by $\exp(-|\vec{\alpha}|^2/2)$
and sum over $l$ we evidently get the normalized multimode CS
$|\vec{\alpha}\ra$ (for any pair $p,\,q$),

\be\lb{ccsexpan}   
|\vec{\alf}\ra =
e^{-\frac{1}{2}|\vec{\alf}|^2}\sum_{l=-\infty}^{\infty}
|\!|\vec{\alf};l,p,q\ra .  \ee
The last equality suggests that the states
$|\!|\vec{\alf};l,p,q\ra $ form overcomplete families in ${\cal H}_l$ for
every $p,\,q$. This is the case: using the overcompleteness
of $|\vec{\alf}\ra$, formula (\ref{ccsexpan}) and the orthogonality relations

\be\label{ortogo1}    
\la p,q,l^\pr;\vec{\alf}|\!|\vec{\alf};l,p,q\ra = 0\quad {\rm for}\quad
l^\pr \neq l,
\ee
one obtains the resolution of unity in ${\cal H}_l$ in terms of the
$u(p,q)$ CS  $|\!|\vec{\alf};l,p,q\ra$,

\be\lb{u(p,q)resolu1}   
  \int d\mu(\vec{\alf})
|\!|\vec{\alf};l,p,q\ra\la p,q,l;\vec{\alf}|\!| = 1_{l}, \quad
d\mu(\vec{\alf}) = \frac{1}{\pi^N}e^{-|\vec{\alf}|^2} d^2\vec{\alf} .
\ee

Now we note that in $u(p,q)$ CS (\ref{u(p,q)bgcssolu}) one complex
parameter, say $\alpha_N$, can be absorbed into the normalization factor by
redefining the rest as

\be\lb{z_i redefine}    
z_1 = \alpha_1\alpha_N, \ld, z_p =
\alpha_p\alpha_N,\,\, z_{p+1} = \alpha_{p+1}/\alpha_N, \dots, z_{N-1} =
\alpha_{N-1}/\alpha_N.
\ee
Then we can write $|\!|\vec{\alf};l,p,q\ra\ = \alf_N^{-l}
|\!|\vec{z};l,p,q\ra$ and

\be\lb{z-u(p,q)bgcs}   
|\!|\vec{z};l,p,q\ra = \sum_{\tld{n}_p-\tld{n}_q=l}
\frac{z_1^{n_1}...z_{N-1}^{N-1}}{\sqrt{n_1! \ld
n_{N-1}!(\tld{n}_p-\tld{n}^\pr_q-l)!}}
|n_1,...,n_{N-1};\tld{n}_p-\tld{n}^\pr_q-l\ra,
\end{equation}
where $\vec{z} = (z_1, \ld, z_{N-1})$. The states
$|\!|\vec{z};l,p,q\ra$ are normalizable in view of

$$1 = \la\vec{\alf}|\vec{\alf}\ra =
e^{-|\vec{\alf}|^2}\sum_{l=-\infty}^{\infty} |\alf_N|^{-2l}\la
q,p,l;\vec{z}|\!|\vec{z};l,p,q\ra, $$
which stems from (\ref{ccsexpan}) and (\ref{ortogo1}). The normalized
states $|\vec{z};l,p,q\ra$ are  $|\vec{z};l,p,q\ra = {\cal
N}\,|\!|\vec{z};l,p,q\ra$, ${\cal N}$ being the normalization constant.

The family $\{|\!|\vec{z};l,p,q\ra\}$ is overcomplete in
${\cal H}_l$ and the resolution of unity reads ($d^2\vec{z}=
\prod_{i}^{N-1}d{\rm Re}z_i\,d{\rm Im}z_i = |\alf_N|^{2(q-1-p)}
\prod_{i}^{N-1}d{\rm Re}\alf_i\,d{\rm Im}\alf_i $)

\be\lb{upqresolu2}   
\bt{l}$1_l = \int d\mu(\vec{z};l,p,q)|\!|\vec{z};l,p,q\ra\la
q,p,l;\vec{z}|\!|,$\\[-2mm]
                     \\
$d\mu(\vec{z},l,p,q) =
F(|\vec{\tld{z}_p}|,|\vec{\tld{z}_q}|;l,p,q)d^2\vec{z}$,
\et \end{equation}
where $|\vec{\tld{z}_p}|^2 = |z_1|^2 + \ld + |z_p|^2$,
\quad $|\vec{\tld{z}_q}|^2 = |z_{p+1}|^2 + \ld + |z_{N-1}|^2$, the
measure weight function being

\be\lb{F}    
F(|\vec{\tld{z}_p}|,|\vec{\tld{z}_q}|;l,p,q) = \frac{1}{\pi^{N}}\int
d^2\alpha_N|\alpha_N|^{2(q-1-p-l)}\exp\l[-\l(\frac{|\vec{\tld{z}_p}|^2}
{|\alf_N|^2} + |\vec{\tld{z}_q}|^2|\alf_N|^2 + |\alf_N|^2\r)\r].
\ee
One can prove that the above measure is unique in the class of smooth
functions of $|z_1|,\ld,|z_{N-1}|$ (see appendix A.3). Thus the
explicit form of $u(p,q)$ BG CS is

\be\lb{z-u(p,q)bgcs2}  
|\vec{z};l,p,q\ra ={\cal N}(|z_1|,\ld |z_{N-1}|;l,p,q)|\!|\vec{z};l,p,q\ra,
\ee
where $|\!|\vec{z};l,p,q\ra$ take the form of superposition
(\ref{z-u(p,q)bgcs}) of multimode Fock states with fixed value $l$ of the
difference number operator $L$, equation (\ref{L}).

Let us note some known particular cases of the $u(p,q)$ BG CS
(\ref{z-u(p,q)bgcs}). Recently the case of $q=1$ and negative $l$, $-l\geq
0$ (then $p = N-1$, $\tld{n}^\pr_q = 0$, $\vec{z}_q = 0$ and
$\vec{z}_p\equiv \vec{z}$) has been considered by Fujii and Funahashi
\ci{Fujii}.  Their resolution unity measure (in ${\cal H}_l$) reads

\be\lb{Fpr}    
d\mu^\pr(\vec{z}) = F^\pr(|\vec{z}|,l,p,1)d^2\vec{z},\quad
F^\pr = \frac{2|\vec{z}|^{-l-p+1}}{\pi^p} K_{-l-p+1}(2|\vec{z}|),
\ee
where $K_\nu(z)$ is the modified Bessel function of the third kind
\ci{Stegun}.
\,\, $F^\pr(|\vec{z}|,l,p,1)$ and $F(|\vec{z}|,l,p,1)$ do not depend on
phases of $z_i$ and are smooth functions of $|z_1|,\ld, |z_{p}|$, i.e.
all order derivatives are finite.  In appendix A.3 we prove that the
resolution unity measures for $u(p,q)$ CS are unique within such a class of
functions, i.e.  $F^\pr(|\vec{z}|,l,p,1)$ and $F(|\vec{z}|,l,p,1)$ should
coincide.  Then using the analyticity property of Bessel functions
$K_\nu(z)$ \ci{Stegun} we establish (proof in appendix A.4) the following
integral representation for $K_\nu(z)$ with $\nu = 0,\pm1,\ld$ and ${\rm
Re}z \geq 0$,

\be\lb{K_nu}   
 K_\nu(2z) = \frac 12 z^{-\nu}\int_{0}^{\infty}dx\,x^{\nu -1}e^{-(x +
z^2/x)}.
\ee
For $p=1,\,q=1$ our states $|\vec{z};l,p,q\ra$ recover (as
the states of \ci{Fujii} do) the BG CS  $|z;k\ra$ for the series  $D^+(k)$
of $su(1,1)$ \ci{BG}, the Bargman index $k$ being expressed in terms of
$l$ as $k = (1 +|l|)/2$.  The irreps with $\pm l$ are equivalent, however
the states $|z;l,1,1\ra$ and $|z;-l,1,1\ra$ are different as one can see
from their definition (\ref{z-u(p,q)bgcs}) (moreover, they are
orthogonal).  Thus our states $|z;\pm l,1,1\ra$ represent two equivalent
but different realizations of BG CS $|z;k\ra$ for $k=(1+|l|)/2
=1/2,1,\ld$. The exact identification is $|\!|z;l\!\leq\! 0,1,1\ra =
|\!|z;k\ra$,\,\, $|\!|z;l\!>\!0,1,1\ra = z^{2k-1}|\!|z;k\ra$.

 The pair CS in quantum optics $|\zeta,q\ra$ \ci{Agar} (defined as
eigenstates of $a_1a_2$ with $a_1^\dg a_1 - a_2^\dg a_2 = q =$ const.)
appear as $u(1,1)$ BG CS $|z;k\ra$ in the two-mode representation ($N=2$
in (\ref{u(p,q)bosrep})).  The identifications is $|\zeta,q\ra =
|z;l,1,1\ra$, i.e. the Agarwal $\zeta$ and $q$ are equal to our $z$ and
$l$ correspondingly.  In view of Eqs.  (\ref{upqresolu2}) - (\ref{Fpr})
the pair CS are overcomplete in the subspaces ${\cal H}_l$.
Our
$|\vec{z};l,p,q\ra$ can be regarded as a generalization of $|\zeta,q\ra$
to the $N$-mode boson system: $|\vec{z};l,p,q\ra$ are invariant under the
annihilation of pairs of two different mode bosons, one from the first $p$
modes, and the other from the last $q$ modes. Note that in the $sp(N,R)$ CS
$|\vec{\alf},C_-,C_+\ra$ there is no such restriction -- these are the
most general states, which are invariant under the annihilation of any
pair of bosons.

The $sp(N,R)$ CS $|\vec{\alf},C_-,C_+\ra$ can be decomposed in terms of
$u(p,q)$ CS $|\vec{z};l,p,q\ra$ with different $l$.  For $N=2$ this
decomposition reads

\bear\lb{g pair cs}  
|\vec{\alf};C_-,C_+\ra = \sum_{l=0}^{\infty}\alf_1^l\tld{C}_l|z;-l,1,1\ra
+ \sum_{l=1}^{\infty}\alf_2^l\tld{C}_l|z;l,1,1\ra,\\
\tld{C}_l =C_+ +
(-1)^lC_-,\quad z=\alf_1\alf_2. \nn
\eear

In analogy to the case of $sp(N,R)$, considered in section 3 we introduce
the $u(p,q)$ multimode squared amplitude cat states
$|\vec{z};l,p,q;D_+,D_-\ra$ as macroscopic superpositions of $u(p,q)$
BG-type CS $|\vec{z};l,p,q\ra$,

\be\lb{u(p,q)cats}  
|\vec{z};l,p,q;D_+,D_-\ra = D_+\,|\vec{z};l,p,q\ra +
D_-\,|-\vec{z};l,p,q\ra,
\ee
which are expected to exhibit squared amplitude squeezing and other
nonclassical properties. In the particular cases of $N=2$, $D_- =
D_+\exp(i\phi)$ the states (\ref{u(p,q)cats}) recover the two-mode
Schr\"odinger cat states, considered recently in \ci{Ger2}.
\vspace{5mm}

\section{Statistical properties of the $N$ mode $sp(N,R)$ BG CS and
their macroscopic superpositions}

In the present section we consider some general statistical properties
of the constructed $sp(N,R)$ algebra related CS and their superpositions
and discuss in greater detail some new subsets of this large family.

 All $sp(N,R)$ BG type CS minimize the Robertson multidimensional
uncertainty relation \ci{Rob} for the Hermitian quadratures
$X_{ij},\,Y_{ij}$ of mutually commuting Weyl lowering operators $E_{ij}$,
since they are eigenstates of all $E_{ij}$ (Proposition 3 of ref.
\ci{Trif97a}),

\be\lb{RUR=}  
\det\sig(\{X_{ij},Y_{ij}\};\vec{\alf},C_-,C_+) = \det
 C(\{X_{ij},Y_{ij}\};\vec{\alf},C_-,C_+),
\ee
where $\sig$ is the matrix of second moments of all observables
$X_{ij},Y_{ij}$ (the uncertainty matrix) and $C$ is the antisymmetric
matrix of all mean commutators of $X_{ij},Y_{ij}$ times $(-i/2)$. The
number of commuting $E_{ij}$ is equal to $(N^2+N)/2$. Robertson
inequality for $n$ observables $X_j$, $j=1,2,\ld,n$, reads

\be\lb{RUR}   
\det\sig(\{X_j\};\Psi) \geq \det C(\{X_j\};\Psi),
\ee
and for a pair of two operators $X,Y$ it reduces to the Schr\"odinger one,
$\Dlt^2X\Dlt^2Y - \sig_{XY}^2 \geq |\la[X,Y]\ra|^2/4$,
where $\sig_{XX} = \la XY+YX\ra/2 -\la X\ra\la Y\ra$ (for greater detail
see for example, \ci{Trif94,Trif97a}). In all $sp(N,R)$ BG CS the
covariances of $X_{ij}$ and $Y_{ij}$ are vanishing, but those of $X_{ij}$
and $X_{kl}$ are not, i.e. the matrix
$\sig(\{X_{ij},Y_{ij}\};\vec{\alf},C_-,C_+)$ is not diagonal.

The subset of Schr\"odinger cats $|\vec{\alf},\varphi\ra$, equation
(\ref{alf;vphi1}), possess several remarkable properties:

(a) They are overcomplete in the whole Hilbert space (see equation
(\ref{alf;vphi resolu}));

(b) The photon statistics in every mode is Poissonian for any $\vec{\alf}$
and $\varphi$.  This follows immediately from the expansion
(\ref{alf;vphi2}) in terms of multimode number states $|\vec{n}\ra =
|n_1,\ld,n_N\ra$;

(c) the states $|\vec{\alf};\varphi\ra$ can exhibit squeezing in the
quadratures $p_j$, $q_j$ (for example, for $|\vec{\alf}|$ close/equal
to $|\alf_i| = 0.5$, $\phi = \pi/4$ and arg$\alf_i$ around $n\pi/2,\, n =
0,1,\ld$, the minimal value of $\Dlt p_i$ and $\Dlt q_i$ being equal to
0.316 -- see the graphics $f_1$ on Fig. 1 );

(d) these states are physically coherent (`true coherent') since they
satisfy the condition of full second-order coherence of the field
\ci{Titulaer}. The latter property again follows from equation
(\ref{alf;vphi2}), which is of the form of generalized CS of Glauber and
Titulaer \ci{Titulaer}.

This interesting subfamily $\{|\vec{\alf};\vphi\ra\}$ of $sp(N,R)$ BG CS
can be generated from the familiar multimode canonical CS by means of the following
operator

\be\lb{S(vphi)}   
S(\vphi) = \exp\l(i(-1)^{\hat{n}}\vphi\r):\quad |\vec{\alf},\varphi\ra =
S(\vphi) |\vec{\alf}\ra.
\ee
where $\hat{n} = a^\dg_1a_1+\ld a^\dg_Na_N$ is the total number operator.
As strange as it may seem $S(\vphi)$ is well defined for any angle
$\vphi$ and is unitary. On any state $|\Psi\ra$ its action is

\be\lb{S(vphi) action}  
S(\vphi)|\Psi\ra = e^{i\vphi}|\!|\Psi\ra_e + e^{-i\vphi}|\!|\Psi\ra_o,
\ee
where $|\!|\Psi\ra_{e,o}$ are the projections of $|\Psi\ra$ on even/odd
subspaces ${\cal H}^{\pm}$. The operator $(-1)^{\hat{n}}$ is Hermitian and
$(-1)^{\hat{n}}\vphi$ may be regarded as a sort of nonlinear multimode
interaction.

In the classification scheme of \ci{Arv1} the (one-mode)
states which possess the above properties (b) and (c) fall into the
subclass of the {\it weakly nonclassical} states. In this scheme the
nonclassical states are subdivided into {\it weakly nonclassical} and {\it
strongly nonclassical} depending on the pointwise nonnegativity or
nonpositivity of the phase averaged ${\cal P}(I)$ Glauber-Sudarshan
diagonal representation $P(\bet)$, $I = |\bet|^2$, $\bet =
\sqrt{I}\exp(i\vtet)$,

\be\lb{cal P}   
{\cal P}(I) = \frac{1}{2\pi}\int_{0}^{2\pi}P(re^{i\vtet})d^2\vtet,\quad
r=\sqrt{I}=|\bet|.
\ee
If ${\cal P}(I) < 0$ for some values of $I$ the state is strongly
nonclassical (then also $P(\bet) < 0$ for some values of $\bet$) and if
${\cal P}(I) \geq 0$ but $P(\bet) \ngeq 0$ the state is said to be weakly
nonclassical \ci{Arv1}. The set of classical states (i.e. $P(\bet)
\geq 0$) is not subdivided. Criteria for phase-insensitive nonclassicality
of single-mode states were also studied in \ci{Klysh}.

The family of $sp(N,R)$ BG CS $|\alf;\varphi\ra$, \,\,equation
(\ref{alf;vphi1}), consists of classical (at $\varphi = 0,\pm\pi/2,\pi$)
and weakly nonclassical states (for $\varphi\neq0,\pm\pi,\pi$) for every
mode since the multimode photon distribution in these states is a
product of one-mode Poisson distributions. There are not strongly
nonclassical states in this family.  Note that at $\varphi =
0,\pm\pi/2,\pi$ the states $|\alf;\varphi\ra$ are the CS $|\alf\ra$ or
$|-\alf\ra$, and at $\varphi = \pi/4$ they coincide with the Yurke--Stoler
states \ci{cats1}.  The states with Gaussian Wigner function are either
classical or strongly nonclassical \ci{Arv1} and strongly nonclassical
states from the latter family all have positive Mandel $Q$ factor
(super-Poissonian statistics) \ci{Mandel,Walls} [$Q = (\Dlt^2\hat{n} -
\la\hat{n}\ra)/\la\hat{n}\ra$, where $\hat{n}= a^\dg a$].

Along these lines we note that in the family of weakly nonclassical
states $|\alf;\varphi\ra$ there are states which exhibit quadrature
squeezing (graphics $f_1$ on figure 1).  Conversely, there exist strongly
nonclassical states (for example, in the family $|\alf,\phi,\psi\ra$,
defined below) which  do not exhibit squeezing of the quadratures of
either $a$ or $a^2$ (nor the $Q$ factor is negative).  Moreover, among
the one-mode $|\alf,\phi,\psi\ra$ there are states with $Q=0$ which are
squeezed or not squeezed, but their photon statistics is not Poissonian
(see graphics on figures 4 and 5). These examples show that $Q=0$ is
not a sufficient condition either for a statistics to be Poissonian or
for a state to be classical.

The quadrature squeezing and/or $Q<0$ are sufficient
conditions for the nonclassicality of the corresponding states [41].
However, they are neither sufficient nor necessary for the strong
nonclassicality as demonstrated below.

In \ci{Arv1} a simple sufficient condition for strong
nonclassicality of the states (i.e. for non positivity of the phase
smeared diagonal $P$ representation ${\cal P}(I)$) is given in terms of
photon number distributions $p_n$,

\be\lb{l_n}   
l_n := (n+1)p_{n-1}p_{n+1} - np_n^2 < 0 \quad {\rm for\,\, some}\,\,\, n>0.
\ee
The distribution $p_n$ is expressed in terms of ${\cal P}(I)$ as \ci{Arv1}

\be\lb{p_n-calP}  
p_n = \int_0^\infty dI{\cal P}(I) p_n{}^{(Pois)}(I),\quad
p_n{}^{(Pois)}(I) = \frac{1}{n!}I^ne^{-I}.
\ee
Distributions $p_n$ which can be represented in the above form with ${\cal
P} \geq 0$ (${\cal P}\ngeq 0$) were recently defined as classical
(nonclassical)\ci{Arv2}. Nonclassicality of $p_n$ means strong
nonclassicality of the corresponding states.

Among $sp(N,R)$ BG CS there are strongly nonclassical states as well (the
definition of strong nonclassicality for multimode states is discussed
below),  such as, for example, the cat states

\be\lb{alf,phi}   
|\vec{\alf},\phi\ra = \tld{\cal N}(|\vec{\alf}\ra +e^{i\phi}|-\vec{\alf}\ra),
\ee
the normalization constant being

$$\tld{\cal N} =\l(2(1+\cos\phi\,e^{-2\tld{r}^2})\r)^{-\frac 12} = \tld{\cal
N}(\tld{r},\phi).$$
In the case of $N=1$ the states (\ref{alf,phi}) have been discussed, for
example, in \ci{Brif96,Arv2} and in the fifth paper of \ci{cats1}.
The probability for totally $n$ photons in $|\vec{\alf},\phi\ra$
(irrespective which mode they belong to), $n=n_1+\ld+n_N$, is found as

\be\lb{p_n(phi)}   
p_n(\tld{r},\phi) =
\tld{N}(\tld{r},\phi)^2e^{-\tld{r}^2}\frac{\tld{r}^{2n}}{n!}s_n(\phi),\quad
s_n(\phi)=2\l(1+(-1)^n\cos\phi\r),\quad \tld{r}=|\vec{\alf}|,
\ee
and the function $l_n(\tld{r},\phi)$ takes the form

\be\lb{l_nphi}    
l_n(\tld{r},\phi) =
 \tld{l}_n(\tld{r},\phi)\l(s_{n-1}(\phi)s_{n+1}(\phi) - s_n^2(\phi)\r),
\ee
where the function  $\tld{l}_n$,

$$\tld{l}_n =
\tld{N}(\tld{r},\phi)^2e^{-\tld{r}^2} \frac{\tld{r}^{4n}}{n!(n-1)!},$$
is nonnegative. The nonnegative factor $s_n(\phi)$  is seen to be a bounded
and oscillating function of both $\phi$ and $n$, $s_n(\phi) =
s_{n+2}(\phi)$. Then for every $\phi\neq \pm \pi/2$ the combination
$s_{n-1}(\phi) s_{n+1}(\phi) - s_n^2(\phi)$ is negative for all $n$ for
which $(-1)^n\cos\phi > 0$. Noting that the total photon number
distribution $p_n(\tld{r},\phi)$, equation (\ref{p_n(phi)}), coincides
with that for the one-mode states $|\tld{\alf},\phi\ra$,
$|\tld{\alf}|=\tld{r}$, we conclude that all one-mode states
$|\alf,\phi\neq\pm\pi/2\ra$ are strongly nonclassical (the strong
nonclassicality of the one-mode states $|\alf,\phi\ra$ was proved also in
the very recent E-print \ci{Arv2}).

One way of generalizing the notion of strong nonclassicality to multimode
states is to apply the above definition to the total photon number
distribution, the other one is to require this for every mode.  One can
easily verify, that in the $N$ mode states $|\vec{\alf},\phi\ra$ the
conditional photon distributions
$p_{n_1,...,n_i,...n_N}(\vec{\alf},\phi)$ for individual mode $i$
($n_{k\neq i}$ being fixed) also obey the inequality (\ref{l_n}). Thus
$|\vec{\alf},\phi\neq\pm\pi/2\ra$ are strongly nonclassical according to
both criteria.

Consider now the {\it squared amplitude quadrature squeezing} \ci{Hill}
in the multimode states.  We first note that the BG-type CS for any Lie
algebra cannot exhibit squeezing of the quadratures $X_{ij}$ and $Y_{ij}$
 of Weyl operators $E_{ij}$ since here the variances of
$X_{ij}$ and $Y_{ij}$ are equal which stems from their eigenvalue property
(\ref{spNRbgcsdef1}) \ci{Trif94}.  In the quadratic boson representation
 $E_{ij} = a_ia_j$ and  $X_{ij}$ (or $Y_{ij}$) squeezing is multimode
squared amplitude squeezing.  The quadrature $X_{ij}$ ($Y_{ij}$) of
 $a_ia_j$ is said squeezed in a state $|\Psi\ra$ if the variance $\Delta
X_{ij}$ ($\Delta Y_{ij}$) is less than its value in the ground state
$|\vec{0}\ra$. Thus quadratic field squeezing does not occur in
$|\vec{\alf};C_-,C_+\ra$. We shall see that macroscopic superpositions of
two such states do exhibit quadratic squeezing. But let us first make
some general remarks about SS of two and several observables.

Squeezing of the two quadratures $X$ and $Y$ of a non-Hermitian
operator $A$ (for definiteness we write $A = X+iY$) can be achieved in
two ways:

(a) in the eigenstates $|z,u,v\ra$ of complex combination $uA + vA^\dg$ (
  generalized intelligent states) \ci{Trif94};

(b) in the eigenstates $|z\ra^{(2)}$ of $A^2$ (generalized cat states).

The first possibility was proved and demonstrated (on the examples of
$SU(1,1)$ and $SU(2)$ generators in series $D^+(k)$ and $D(j)$) in
\ci{Trif94,Trif96a}. These SS minimize the Schr\"odinger inequality and
therefore were called Schr\"odinger (or generalized) IS. A particular
case of SS of type (a) are the SS for general systems \ci{NT},
introduced as states minimizing the Heisenberg inequality, which
is a particular case of that of Schr\"odinger. The second possibility (b)
can be easily proved by calculations using the eigenvalue condition of
$A^2$ and taking into account the Schr\"odinger relation.  This can also
be checked directly on the example of the following two type of
superposition states

\be\lb{type(b)SS}   
|z;\vphi\ra = \cos\vphi|z\ra + i\sin\vphi|-z\ra,\quad
|z,\phi\ra = {\cal N}(|z\ra + e^{i\phi}|-z\ra),
\ee
where $|z\ra$ are eigenstates of $A$, $A|\pm z\ra = \pm z|\pm z\ra$. These
states can exhibit squeezing according to the stronger criterion, given in
\ci{Trif94} (see also below).  \,\, $|z;\vphi\ra$ and $|z,\phi\ra$ are
eigenstates of $A^2$ (and not of $A$, unless $\vphi = n\pi/2$,
$n=0,1,\ld$).  Eigenstates $|z\ra^{(2)}$ of $A^2$ which are not
eigenstates of $A$ are superpositions of $|\pm z\ra$.  Therefore SS of
type (b) are cat states. \,$|z;\vphi\ra$ and $|z,\phi\ra$ in
(\ref{type(b)SS}) are examples of such SS for any $A$ for which
eigenstates $|\pm z\ra$ do exist. In fact first- and higher order squeezing
of $X$ and $Y$ can occur in states which are eigenvectors of $A^n$ for any
$n\geq 1$ and such eigenvectors can be easily expressed as discrete
superpositions of several $|z\ra$.

The operator $S(u,v)$ which transform the nonsqueezed $|z\ra$ to the
SS of type (a), $|z,u,v\ra$, was defined in \ci{Trif94,Trif97a} as
generalized squeeze operator (if $A=a$ then $S(u,v)$ is the known
canonical squeeze operator \ci{Walls,XMa}). Having established that
eigenstates $|z\ra^{(2)}$ of $A^2$ can universally exhibit squeezing of
the quadrature of $A$ we can define, in analogy to the previous case,  a
{\it second kind of squeeze operator} $S_{II}$ by means of the
relation

\be\lb{S_II} 
|z\ra^{(2)} = S_{II}|z\ra .
\ee
We can point out an example of such a squeeze operator - that is the
operator $S(\vphi)$ of equation (\ref{S(vphi)}). It maps the multimode
CS $|\vec{\alf}\ra$ to the weakly nonclassical states ($sp(N,R)$ BG CS)
$|\vec{\alf};\vphi\ra$ which are eigenstates of $a_ia_j$ (the $sp(N,R)$ BG
CS) and do exhibit quadrature squeezing (see the graphics $f_1$ on figure
1).

The main difference between the above two types of SS is the following: SS
of type (a) can exhibit arbitrary strong squeezing of $X$ or $Y$, while the
squeezing in SS of type (b) is always bounded, since eigenstates of $A^2
\sim (X+iY)^2$ can never tend to an eigenstate of $X$ or $Y$. Family of
states in which arbitrary strong squeezing (`ideal squeezing') of $X$ or
$Y$ is possible could be called {\it ideal $X$-$Y$ SS}. So the
Schr\"odinger IS, in particular the canonical SS \ci{Walls}, are ideal
$p$-$q$ SS. We follow the definition of $X$-$Y$ SS according to ref.
\ci{Trif94}:  a state $|\Psi\ra$ is $X$-$Y$ SS if $\Dlt X < \Dlt_0$ or
$\Dlt Y < \Dlt_0$, where $\Dlt_0$ is the lowest level at which the
equality $\Dlt X = \Dlt Y$ can be maintained. The lowest level is reached
on some eigenstate $|z_0\ra$ of $A$. For the quadratures of $a^k$,
k=1,2,\ld, and $(a_ia_j)^k$ the lowest level is reached in the ground
state $|0\ra$.  Linear and/or quadratic squeezing in ideal one-mode
squared amplitude SS (eigenstates $|z;u,v\ra$ of $ua^2+va^2$) is
considered in papers (for different ranges of parameters $u,\,v$)
\ci{Hill,Trif95,Trif96b,Brif97a,Brif97b,Marian}.  The diagonalization of
$ua^k + va^{\dg k}$ for $k>2$ is discussed in the very recent E-print
\ci{Nagel}.

A family of states in which squeezing of quadratures of any product
$a_ia_j$, $i,j = 1,2,\ld,N$, can occur should be called a family of {\it
multimode squared amplitude SS}. Example of such multimode SS is given by
the Robertson IS \ci{Trif97a}, which should be eigenstates of complex
combinations $u_{kl;ij}a_ia_j + v_{kl;ij}a^\dg_ia^\dg_j$ (summation over
repeated indices).  These are {\it ideal} multimode squared amplitude SS.
Multimode quadratic SS of type (b) are defined as eigenstates of all
squared products $(a_ia_j)^2$. They take the form (\ref{SA cats}).\\[0mm]

Next we consider cat type squared amplitude  SS.  One example of two-mode
cat-type second-order SS is considered in paper \ci{Ger2}, which, however,
was examined for ordinary squeezing only.  Here we provide examples of
multimode cat-type SS which can exhibit both quadratic and linear
squeezing and other interesting statistical properties.  Such SS are the
following macroscopic superpositions $|\vec{\alf},\phi,\psi\ra$ of the
$sp(N,R)$ algebraic CS $|\vec{\alf},\phi\ra$ ($|\vec{\alf},\phi\ra$ are
defined in equation (\ref{alf,phi})):

\be\lb{alf,phi,psi}   
|\vec{\alf},\phi,\psi\ra = {\cal N}(|\vec{\alf},\phi\ra +
e^{i\psi}|-\vec{\alf},\phi\ra),
\ee
where ${\cal N}$ is the normalization constant, which obey
(\ref{normcond1}) and has the form
\be\lb{calN}    
{\cal N}(\tld{r},\phi,\psi)= \frac{1}{\sqrt{2}}\l(1 + 2
\tld{\cal N}^2\,e^{-\tld{r}^2}\l(\cos\phi\,\cos(\tld{r}^2-\phi+\psi) +
\cos(\tld{r}^2+\phi-\psi)\r)\r)^{-\frac 12},
\ee
$\tld{\cal N}$ being given in equation (\ref{calN}) and $\tld{r} =
|\vec{\alf}|= \sqrt{|\alf_1|^2 + |\alf_2|^2 +\ld+|\alf_N|^2}$.

We demonstrate the quadrature squeezing on the example of individual mode
operators $a_i$ (linear squeezing) and $a_i^2$ (quadratic squeezing). Note
that $|\vec{\alf},\phi,\psi\ra$ are not factorized over the different
modes.  The variances $\Dlt p_i$ and $\Dlt q_i$ of the quadratures of
$i$ mode annihilation operator $a_i$, $a_i = (q_i + i p_i)/\sqrt{2}$ are

\bear\lb{Dlt q,p}   
\Dlt^2p_i(\tld{r},r_i,\tet_i,\phi,\psi) = \frac12 + \la a_i^\dg a_i\ra
- {\rm Re}\la a_i^2\ra - 2({\rm Im}\la a_i\ra)^2,\nn \\
\Dlt^2q_i(\tld{r},r_i,\tet_i,\phi,\psi) = \frac 12 + \la a_i^\dg a_i\ra
+ {\rm Re}\la a_i^2\ra - 2({\rm Re}\la a_i\ra)^2,
\eear
where $r_i = |\alf_i|$, $\tet_i = \arg\alf_i$ and

\bear\lb{a1,adga}   
\la a_i\ra = -2\alf_i{\cal N}^2\tld{\cal N}^2e^{-\tld{r}^2}\sin\phi\,
(1+i)\l(e^{-\tld{r}^2} + \cos(\tld{r}^2\!-\!\phi\!+\!\psi) +
\sin(\tld{r}^2\!-\!\phi\!+\!\psi)\r),\nn\\
\la a_i^\dg a_i\ra = 4r_i^2{\cal N}^2\,\tld{\cal
N}^2\l(1-\cos\phi\,e^{-2\tld{r}^2}
- e^{-\tld{r}^2}\l(\cos\phi\,\sin(\tld{r}^2\!-\!\phi\!+\!\psi) +
\sin(\tld{r}^2\!+\!\phi\!-\!\psi)\r)\r).
\eear
As functions of $\tet_i$ the variances of $p_i$ and $q_i$ oscillate with
period $\pi$ and $\Dlt p_i(\tet_i+\pi/2) = \Dlt q_i(\tet_i)$.
Linear squeezing is exhibited in states, for example,
$|\vec{\alf},0,\psi\ra$ with $r_i = 0.05$, $\tld{r}$ close to $r_i$,
$\theta_i = \pi/4$, $\phi = 0$ and $\psi$ around $3.152$ (see figure 1).
Maximal $p_i$ ($q_i$) squeezing is obtained when $\tld{r} = r_i$ (i.e.
when only one mode is excited). Here $\Dlt^2 p_i \geq 0.275 = \Dlt^2
p_i(0.05,0.05,\pi/4,0,3.131) = \Dlt^2 q_i(0.05,0.05,\pi/4,0,3.153)$. When
$\tld{r}$ is increasing the graphics of $\Dlt p_i$ and $\Dlt q_i$ (as
functions of the angles $\phi$ and $\psi$) become smoother and tend to a
constant value, independent of the superposition parameters $\phi$ and
$\psi$. In the above states $|\vec{\alf},0,\psi\ra$ the Mandel factor
$Q_i$ for the mode $i$ is negative in the vicinity of $\psi = \pi$ only.
By its definition the quantity $\tld{r}^2$ coincides with the intensity of
the field (the total mean number of photons $\la
\vec{a}^\dg\vec{a}\ra = \sum_i^N\la a_i^\dg a_i\ra$) in the multimode CS
$|\vec{\alf}\ra$. The field intensity in the multimode superposition
states $|\vec{\alf},\phi,\psi\ra$ reads

\be\lb{<n_tot>}   
\la\vec{a}^\dg\vec{a}\ra =
4\tld{r}^2 {\cal N}^2\tld{\cal N}^2\l(1\!-\!\cos\phi\,e^{-2\tld{r}^2}\!-\!
e^{-\tld{r}^2}\l(\cos\phi\,\sin(\tld{r}^2\!-\!\phi\!+\!\psi) +
\sin(\tld{r}^2\!+\!\phi\!-\!\psi)\r)\r).
\ee
We see that $\la\vec{a}^\dg\vec{a}\ra$ is an increasing function of
$\tld{r}$, $\tld{r} =|\vec{\alf}|$.

The variances of the quadratures $X_i,\,Y_i$ of the squared
individual mode amplitude $a_i^2$, $a_i^2 = (X_i + i Y_i)/\sqrt{2}$, in
$|\vec{\alf},\phi,\psi\ra$ is easily obtained in the form

\bear\lb{Dlt X,Y}    
\Dlt^2X_i(\tld{r},r_i,\tet_i,\phi,\psi) = 1 + 2\la a_i^\dg a_i\ra + \la
a_i^{\dg 2}a_i^2\ra + r_i^4\cos(4\tet_i) - 2({\rm Re}\la
a_i^2\ra)^2,\nn \\
\Dlt^2Y_i(\tld{r},r_i,\tet_i,\phi,\psi) = 1 + 2\la a_i^\dg a_i\ra + \la
a_i^{\dg 2}a_i^2\ra - r_i^4\sin(4\tet_i) - 2({\rm Im}\la a_i^2\ra)^2.
\eear
where

\bear\lb{a2}     
\la a_i^2\ra = -4i\alf_i^2{\cal N}^2\tld{\cal N}^2e^{-\tld{r}^2}
\l(\cos\phi\sin(\tld{r}^2-\phi + \psi) - \sin(\tld{r}^2+\phi-\psi)\r),\nn
\\ \la a_i^{\dg 2}a_i^2\ra = 2r_i^4{\cal N}^2\, \l(1-2\tld{\cal
N}^2e^{-\tld{r}^2}\l(\cos\phi\cos(\tld{r}^2-\phi+\psi) + \cos(\tld{r}^2 +
\phi-\psi)\r)\r).
\eear
As functions of the angle $\tet_i$ the variances
of $X_i$ and $Y_i$ oscillate with period $\pi/2$ and $\Dlt
X_i(\tet_i+\pi/4) = \Dlt Y_i(\tet_i)$.  \vspace{2mm}

\bc
\bt{ll}
\hs{-1.2cm}\input{fg1a.pic}     &\hs{-1cm} \input{fg2a.pic}\\
                                  &        \\
\hs{-5mm}{\sm Fig. 1. Amplitude quadrature squeezing in }
& {\sm Fig. 2. Squared amplitude quadrature squeezing } \\

\hs{-5mm}{\sm the $sp(N,R)$ BG CS $|\vec{\alf};\vphi\ra$ and the
superpo-} & {\sm in superpositions states $|\vec{\alf},\phi,\psi\ra$} for
$r_i=0.8$.
\\
\hs{-5mm}{\sm sitions $|\vec{\alf},\phi,\psi\ra$, equation (61), for
$r_i\!=\!|\alf_i|\!=\!0.05$.}  & {\sm
$\Dlt\tld{X_i}=\Dlt\tld{X_i}(\tld{r},r_i,\tet_i,\phi,\psi)$,
$\tld{X}_i^2\!=\!X_i$ or $Y_i$, $\tld{r}=|\vec{\alf}|$}.
\\
\hs{-5mm}{\sm $\Dlt\tld{q_i}=\Dlt\tld{q_i}(\tld{r},r_i,\tet_i,\phi,\psi)$,
$\tld{r}\!=\!|\vec{\alf}|$, $\tld{q}_i\!=\!q_i$ or $p_i$. } & {\sm
$g_1\!=\!\Dlt^2X_i(r_i,r_i,\frac{\pi}{4},0,\psi)\!=\!\Dlt^2Y_i
(r_i,r_i,-\frac{\pi}{4},0,\psi)$,}
\\
\hs{-5mm}{\sm $f_1\!=\!\Dlt^2q_i(r_i,r_i$,-$\frac{\pi}{2},\vphi)$,\,
$f_2\!=\Dlt^2p_i(r_i,r_i,\frac{\pi}{4},0,\psi)$, }    & {\sm
$g_2\!=\!\Dlt^2X_i(1,r_i,\frac{\pi}{4},0,\psi)$,
$g_3\!=\!\Dlt^2X_i(1.2,r_i,\frac{\pi}{4},0,\psi),$}
\\
\hs{-5mm}{\sm $f_3\!=\!
\Dlt^2q_i(r_i,r_i,\frac{\pi}{4},0,\psi)$,\,\,$f_4\! =\! \Dlt^2
p_i(4r_i,r_i,\frac{\pi}{4},0,\psi)$, }  & {\sm
$g_4\!=\!2\Dlt^2p_i(r_i,r_i,\frac{\pi}{4},0,\psi)\!=
\!2\Dlt^2q_i(r_i,r_i,-\frac{\pi}{4},0,\psi)$.}
\\
\hs{-5mm}{\sm $|\vec{\alf},\vphi\ra$ are weakly nonclassical states for
eve-}      & {\sm  Joint $X$ and $p$ (or $Y$ and $q$) squeezing occurs in}
\\
\hs{-5mm}{\sm ry mode,\,\, $|\vec{\alf},\phi,\psi\ra$ are strongly
nonclassical.} &{\sm the interval $ 6.4 <\psi< 7.4 $. } \et \ec \vs{8mm}

\bc
\bt{ll}
\hs{-1cm}\input{fg3a.pic}  & \hs{-0.3cm} \input{fg4a.pic}\\
                  &       \\[6mm]
\makebox(60,70)[l]{
\begin{minipage}{7cm}
{\small Fig. 3. Probabilities $p_n(\tld{r},\phi,\psi)$ to find $n$ photons
in the multimode superposition states $|\vec{\alf},\phi,\psi\ra$ as
functions of $\tld{r}=|\vec{\alf}|$ for different values of $\phi$ and
$\psi$.
$p_0=p_0(\tld{r},\frac{\pi}{2},\pi)$,\,
$p_1=p_1(\tld{r},\pi,-\frac{\pi}{2})$,\,
$p_2=p_2(\tld{r},0,\pi)$,\,
$p_3=p_3(\tld{r},\pi,\frac{\pi}{2})$,\,
$p_4=p_4(\tld{r},\frac{\pi}{4},\frac{\pi}{4})$,\,
$p_5=p_5(\tld{r},\pi,-\frac{\pi}{2})$. }
\end{minipage} }      &

$\quad$\makebox(60,70)[l]{
\begin{minipage}{7cm}
{\small Fig. 4. Oscillating photon number distributions
$p_n(\tld{r},\phi,\psi)$ in strongly nonclassical states
$|\vec{\alf},\phi,\psi\ra$ for different values of $\tld{r}$, $\phi$ and
$\psi$. $p_n1=p_n(0.8,0,7.3)$ ($Q>0$, $\Dlt p\geq 0.38$, $\Dlt X\geq 0.73$),\,
$p_n2=p_n(2.2,\pi,-\frac{\pi}{2})$ ($Q<0$),\, $p_n3=p_n(0.55,5.0914,0)$
($Q=-0$, $\Dlt p\geq 0.38$, $\Dlt X>1$). }
\end{minipage} }
\et
\ec \vs{0.3cm}

The variances $\Dlt X_i$ and $\Dlt Y_i$ are squeezed if
they are less than their value of $1$ in the ground state $|0\ra$. This
holds, for example, in states $|\vec{\alf},0,\psi\ra$ with $\tld{r}$
close/equal to $r_i \leq 1$, $\theta_i= n\pi/4$, $\phi =0$ and $\psi$
around zero, the minimal value of $\Dlt^2 X_i$ and $\Dlt^2 Y_i$ (at
$\tld{r}=r_i=0.88$) being equal to 0.69.
In the states $|0.8e^{\pm i\pi/4},0,\psi\ra$ linear and quadratic
squeezing can occur simultaneously (see graphics $g_1$ and $g_4$ on figure
2: joint $X_i$- and $p_i$- ($Y_i$- and $q_i$-) squeezing occurs in the
interval $6.4 \leq \psi \leq 7.4$). In the above interval $Q_i >0$, where
$Q_i$ is the Mandel factor for the individual mode $i$.

On figure 2 graphics are shown of $\Dlt^2 X_i(\tld{r},r_i,\pi/4,0,\psi)$ as
a function of the angle $\psi$ for fixed $r_i = 0.8$, and three
different values of the total excitation parameter $\tld{r}$, $\tld{r} =
0.8 = r_i$ (i.e. only mode $i$ excited, graphics $g_1$), $\tld{r} = 1$
(graphics $g_2$) and $\tld{r} = 1.2$ (graphics $g_3$). One sees, that
graphics of $\Dlt X_i(\psi)$ become rapidly smoother and tend to a
constant value when $\tld{r}$ is increasing.

An important statistical property of all states $|\vec{\alf},\phi,\psi\ra$
is that they are {\it strongly nonclassical} in the sense of definition of
\ci{Arv1} (discussed above), which we apply here to the total
photon number (and to the conditional individual mode number) distribution
in the multimode states.  The total photon number distribution
$p_n(\tld{r},\phi,\psi)$ takes the form similar to that of equation
(\ref{p_n(phi)}),

\bear\lb{p_n(phi,psi}   
p_n(\tld{r},\phi,\psi) = \tld{p}_n(\tld{r},\phi,\psi)\,s_n(\phi,\psi),\nn\\
s_n(\phi,\psi) = \l|1+(-1)^ne^{i\phi} + i^n e^{i\psi} +
(-i)^ne^{i(\psi-\phi)}\r|^2,
\eear
where
$$\tld{p}_n(\tld{r},\phi,\psi) = {\cal N}^2(\tld{r},\phi,\psi)\tld{\cal
N}^2(\tld{r},\phi)\,e^{-\tld{r}^2}\,\frac{\tld{r}^{2n}}{n!}.
$$
The factor $s_n(\phi,\psi)$ is bounded from above and as a function on $n$
oscillates with period $4$. Therefore the inequality (\ref{l_n}) is
satisfied in all states $|\vec{\alf},\phi,\psi\ra$ for those $n$ for
which $s_n$ reaches its local maximum. This proves that
all $|\vec{\alf},\phi,\psi\ra$ are strongly nonclassical. Note that the
factors $l_{n_i}$ for conditional distribution of $n_i$ ($n_{k\neq i}$
fixed) also satisfy the inequality (\ref{l_n}).
\vs{5mm}

\bt{ll}
\hs{-1.3cm}\input{fg5a.pic}
& \hs{1mm}\makebox(6.5,8)[l]{\vs{5cm}
\begin{minipage}{7.2cm}
{\small
Fig. 5. Nonoscillating photon number distributions in strongly
nonclassical states $|\vec{\alf},\phi,\psi\ra$  for different values of
$\tld{r}=|\vec{\alf}|$, $\phi$ and $\psi$.\,\, $p_n1 = p_n(0.55,2.246,0)$
\, ($Q<0$, $\Dlt q\geq 0.5$, $\Dlt X\geq 1$),\,\, $p_n2 =
p_n(0.55,2.33,0)$\, ($Q<0$,
$\Dlt q\geq 0.43$, $\Dlt X\geq 1$),\,\, $p_n3
= p_n(0.55,2.234384,0)$\, ($Q= +0$, $\Dlt q\geq 0.5$, $\Dlt X\geq 1$),\,\,
$p_n4 =$ Poisson distribution with $\la a^\dagger a\ra = 0.685$ as in $p_n3$.}
\end{minipage}
}\et \vs{5mm}

As in the case of $|\vec{\alf},\phi\ra$ here again $p_n(\tld{r},\phi,\psi)$
coincides with the probability to find $n$ photons in the one-mode states
$|\tld{\alf},\phi,\psi\ra$, $|\tld{\alf}| = \tld{r}$, $\tld{\alf} =
\tld{r}e^{i\tld{\tet}}$.  The distributions $p_n(\tld{r},\phi,\psi)$ does
not depend on $\tld{\tet}$. It can be oscillating or nonoscillating and
with positive, negative or vanishing individual mode $Q$ factor. No
definite relations exist between the sign of $Q$, the photon number
oscillations and the amplitude quadrature squeezing: all possible
combinations of these three properties can be found in strongly
nonclassical states $|\vec{\alf},\phi,\psi\ra$.  In figures 4 and 5
representative graphics of oscillating (figure 4) and nonoscillating
(figure 5) photon distributions are shown. The sign of the corresponding
$Q$ and the inequalities for $\Dlt q(\tld{\tet})$ and $\Dlt X(\tld{\tet})$
for each graphics are also given. In the recent E-print \ci{Arv2} examples
of classical states with oscillating photon distributions were pointed
out.  Thus photon number oscillations are neither necessary nor sufficient
for nonclassicality of quantum states.

The $Q$ factor is bounded, $Q\geq -1$, and when $Q=-1$ then the variance
$\Dlt n$ of $\hat{n}$ is vanishing.  This means \ci{Trif94} that the
corresponding state is an eigenstate $|n\ra$ of $\hat{n}$ (a Fock state)
and $p_n(|n\ra) = 1$. In figure 3 photon probabilities
$p_n(\tld{r},\phi,\psi)$, $n=0,1,2,3$,  are shown as functions of
$\tld{r}$ for several values of $\phi$ and $\psi$. At $\tld{r}\rar 0$ one
obtains $p_n = 1$. For $N>1$ this yields the {\it finite superpositions of
multimode Fock states} $|\vec{n}\ra$, $n_0+\ld+n_N=n$,  and for the one
mode case, $N=1$, -- the number state $|n\ra$ with $n=0,1,2$ or $n=3$. We
see from figure 3 that practically the states $|\vec{\alf},\phi,\psi\ra$
with the corresponding $\phi,\,\psi$ coincide with Fock states $|n\ra$,
$n=1,2,3$, for $|\alf| \leq 0.5$ (then $p_n > 0.99995$). In multimode case
for $|\vec{\alf}| \leq 0.5$ the specific form of superposition of several
$|\vec{n}\ra$ depends on the specific values of $|\alf_i|$,
$|\alf_1|+\ld+|\alf_N| \leq 0.5$. If $\alf_{k\neq i}=0$ then all $n$
photons/bosons are in the mode $i$, i.e. the Fock state is $|\vec{n}\ra =
|0,\ld,n_i\!=\!n,0,\ld,0\ra$.  There is a growing interest in
obtaining Fock states from macroscopic superpositions of (so far mainly
one mode) CS $|\alf\ra$ (see \ci{Jans} and references therein). Here we
provided example, probably the first one, of obtaining Fock states of
multimode systems.
\vspace{5mm}

\section{Concluding remarks}

We have constructed and discussed some properties of $sp(N,R)$ and $u(p,q)$
algebraic (algebra related) CS in the quadratic boson
representation.  These states are a generalization of the $su(1,1)$ CS of
Barut and Girardello \ci{BG} and are constructed as eigenstates of all
mutually commuting Weyl lowering operators.  The quadratic boson
realizations of $sp(N,R)$ and $u(p,q)$ are reducible. Therefore the
corresponding group related CS \ci{KlSk} are not overcomplete in the whole
Hilbert space of states ${\cal H}$. The BG-type CS are very large sets
and afford the possibility to resolve the unity operator in ${\cal H}$ by
means of some subsets. We pointed out such subsets of the $sp(N,R)$
algebra related CS (and their superpositions as well) and wrote down the
relations between the established $u(p,q)$ CS representations and the
familiar $N$ mode canonical CS representation, in particular between the
$su(1,1)$ BG CS and the two-mode canonical CS representations.

The new states can exhibit interesting statistical properties, such as
amplitude quadrature squeezing, sub- and super-Poissonian photon
statistics and oscillations in photon number distributions. All states
from the overcomplete subfamily $|\vec{\alf};\varphi\ra$ of the $sp(N,R)$
BG-type CS are weakly nonclassical \ci{Arv1} and (some of them) can exhibit
amplitude quadrature squeezing as well. Strongly nonclassical \ci{Arv1}
$sp(N,R)$ algebra related CS were also pointed out.

Noting that the BG type CS $|\vec{z}\ra$ can not exhibit squeezing of the
quadratures of Weyl generators $E_{ij}$ we anticipated that such squeezing
should occur in eigenstates of $E^m_{ij}$, $m\geq 2$, which for
$E^2_{ij} = (a_ia_j)^2$ are called multimode squared amplitude
Schr\"odinger cat states.  Squared amplitude squeezing in the individual
modes is demonstrated in the superpositions $|\vec{\alf},\phi,\psi\ra$ of
two $sp(N,R)$ CS.  These are strongly nonclassical states and at small
$|\vec{\alf}|$ ($|\vec{\alf}|<0.5$) and for specific values of the angles
$\phi,\,\psi$  practically coincide with superpositions of several
multimode Fock states $|\vec{n}\ra$ with total number of photons/bosons
$n=1,2$ or $n=3$. If $\alf_{k\neq i} = 0$ then all $n$ photons are of the
mode $i$, i.e. we have a single multimode Fock state. The Fock state
engineering via discrete superpositions of canonical CS $|\alf\ra$ is of
current interest in the literature (one mode mainly). We have shown that
discrete superpositions of multimode canonical CS are naturally
encompassed in the framework of $sp(n,R)$ BG-type CS and their linear
combinations. The weakly nonclassical $sp(N,R)$ BG CS
$|\vec{\alf};\vphi\ra$ can be generated from CS $|\alf\ra$ by means of the
second kind (unitary) squeeze operator.  In greater detail this should be
considered elsewhere.  \vs{3mm}

{\bf Acknowledgment}. The work was partially supported by Bulgarian
Science Foundation under contract no F--559.
\vspace{5mm}

\section{Appendix}

\subsection{A.1. The correspondence rule between the $N$ mode canonical CS
and $u(p,q)$ BG-type CS representations}

The multimode CS $|\vec{\alf}\ra$ are overcomplete in the $N$ mode boson
system Hilbert space ${\cal H}$, spanned by the number states $|\vec{n}\ra =
|n_1,\ld,n_N\ra$. In canonical CS representation a state $|\Psi\ra$ is
represented by analytic function $F_{CCS}(\vec{\alf};\Psi)$ of $N$
variables $\alf_i$, $i=1,\ld,N$,

\be\lb{F_CCS N}   
F_{CCS}(\vec{\alf},\Psi) = \la\vec{\alf}^*|\!|\Psi\ra, \quad
 |\!|\vec{\alf}\ra = \sum_{n_1,\ld,n_N}\frac{\alf_1^{n_1}\ld\alf_N^{n_N}}
{\sqrt{n_1!\ld n_N!}}|\vec{n}\ra \nn
\ee

The $u(p,q)$ BG-type CS $|\!|\vec{z};l,p,q\ra$, equation
(\ref{z-u(p,q)bgcs}), are overcomplete in subspaces ${\cal H}_l$,
satisfying the resolution unity equation (\ref{upqresolu2}). Hereafter if
a state $|\Psi\ra \in {\cal H}_l \subset {\cal H}$, then in the $u(p,q)$
CS representation this state is represented by the analytic functions of
$N-1$ variables $z_k$, $k=1,\ld,N-1$,

\be\lb{F_l}    
F_{l}(p,q,\vec{z};\Psi) = \la q,p,l;\vec{z}^*|\!|\Psi\ra.
\ee
The relation between the two representatives of $|\Psi\ra$ immediately
follows from the expansion (\ref{ccsexpan}) and equations (\ref{z_i
redefine}) and (\ref{z-u(p,q)bgcs2}):

\be\lb{F_l-CCS}  
F_{CCS}(\vec{\alf},\Psi) = \sum_{l=-\infty}^{\infty}
\alf_N^{-l}F_l(p,q,\vec{z};\Psi), \quad z_k \,\,\mbox{given by equation
(\ref{z_i redefine})}.  \ee This formula is efficient for the transition
from $\{F_l\}$ to $F_{CCS}$ if one knows the representatives
$F_l(p,q,\vec{z};\Psi)$. In the opposite direction the transition formula
is easily obtained from (\ref{F_l}), (\ref{ccsexpan}) and the
orthogonality between ${\cal H}_l$ and ${\cal H}_{l^\pr\neq l}$,

\be\lb{CCS-F_l}  
F_{l}(p,q,\vec{z}^\pr;\Psi) = \frac{1}{\pi^N}\int d^2\vec{\alf}\alf_N^l \la
q,p,l;\vec{z}^{\pr *}|\!|\vec{z}(\vec{\alf});l,p,q\ra e^{-|\vec{\alf}|^2}
F_{CCS}(\alf;\Psi),
\ee
where $\vec{z}(\vec{\alf})$ is given according to (\ref{z_i redefine}) and
$|\!|\vec{z};l,p,q\ra$ is the state (\ref{z-u(p,q)bgcs}).

The $u(p,q)$ CS representation is yet not fully specified (this could be a
subject of separate work), except for the case of $p=1=q$ ($N=2$ when it
coincides with the well known $su(1,1)$ BG CS representation \ci{BG}. In
this case the relation (\ref{F_l-CCS}) is rewritten in the simpler form

\be\lb{BGl-CCS}  
F_{CCS}(\alf_1,\alf_2,\Psi) = F_{l=0}(z;\Psi) +
\sum_{l=1}^{\infty} (\alf_1^{l} + \alf_2^{l})F_{l}(z;\Psi),\quad z =
\alf_1\alf_2.
\ee
The BG representation is given \ci{BG} in terms of Bargman index $k$, not
in terms of $l$: $|\Psi\ra \rar F_{BG}(z,k;\Psi)$. The relation between
$l$ and $k$ is $l = \pm \sqrt{4k(k-1)+1}$, or

\be\lb{k-l}    
k=\frac 12 (1 + |l|), \quad l = n_1 - n_2,
\ee
One has $F_{l\leq 0}(z;\Psi) = F_{BG}(z,k\!=\!(1\!+\!|l|)/2;\Psi)$,\,
$F_{l>0}(z;\Psi) = F_{BG}(z,k\!=\!(1\!+\!l)/2;\Psi)$ and

\be\lb{BGk-CCS}  
F_{CCS}(\alf_1,\alf_2;\Psi) = F_{BG}(z,k\!=\!1/2;\Psi) +
\sum_{k\geq 1}(\alf_1^{2k-1} + \alf_2^{2k-1})F_{BG}(z,k;\Psi),\quad z =
\alf_1\alf_2.
\ee
The relation (\ref{k-l}) stems from the definition of $k$ by means of the
Casimir operator: $C_2 = K_3^2 -\frac{1}{2}(K_+K_- + K_-K_+)  = k(k-1)$. In
the two-mode $su(1,1)$ representation (\ref{K_i N=2}) we have $C_2 = -1/4 +
L^2/4$ which tell us that both $l$ and $-l$ lead to the same values of
$C_2$, that is the representations  realized in the subspaces with $\pm l$
are equivalent.  However for the transitions between canonical CS and BG
representations the sign of $l$ is significant and it is taken
into account in (\ref{BGl-CCS}) by the identification of the standard
$su(1,1)$ notation $|n+k,k\ra$ of the eigenstates of $K_3$ once with
the two-mode Fock state $|n+|l|,n\ra$ (the first term in the sum in
(\ref{BGk-CCS}) and second with $|n,n+|l|\ra$ (the second term in the sum
in (\ref{BGk-CCS}).  Keeping in mind the latter identification rule we can
express the two-mode canonical CS $|\alf_1,\alf_2\ra$ in terms of BG CS
$|z;k\ra$,

\be\lb{|BGk>-|CCS>}  
|\alf_1,\alf_2\ra = |z;k\!=\!1/2\ra  +
\sum_{k\geq 1}(\alf_1^{2k-1} + \alf_2^{2k-1})|z;k\ra,\quad z =
\alf_1\alf_2.
\ee

\subsection{A.2. On the overcompleteness of eigenstates of $A^{2^n}$}

Let $\{|\vec{z}\ra\}$ be an overcomplete family of eigenstates of a $N$
non-Hermitian operator $A_i$, $A_i|\vec{z}\ra = z_i|z\ra$,

\be\lb{|vec{z}>1resolu}  
 1 = \int d\mu(\vec{z},\vec{z}^*)|\vec{z}\ra\la \vec{z}|,
 \ee
where $d\mu(\vec{z},\vec{z}^*) = F(|z_1|,\ld,|z_N|)d^2\vec{z}$. We note
the requirement the weight function $F$ not to depends on the phases of
$z_i$. This holds for the $N$ modes canonical CS, $sp(N,R)$ and $u(p,q)$ CS.
Consider the sequence of families $|\vec{z}\,^{2^n};\vphi_1,\ld,
\vphi_n\ra$,

\be\lb{|z;vecvphi>}  
|\vec{z}\,^{2^n};\vphi_1,\ld,\vphi_n\ra =
\cos\vphi_n|\vec{z}\,^{2^{n-1}};\vphi_1,\ld,\vphi_{n-1}\ra + i\sin\vphi_n
|-\vec{z}\,^{2^{n-1}};\vphi_1,\ld,\vphi_{n-1}\ra,
\ee
where $\vphi_k$, $k =1,\ld,n$, are angle parameters, $n$ is any positive
integer and $\vec{z}\,^{2^n}$ is the $N$ component column
$(z_1^{2^n},\ld,z_N^{2^n})$ of eigenvalues of powers $A_i^{2^{n}}$ of
operators $A_i$,

\be   
A_i^{2^{n}}|\pm \vec{z}\,^{2^n};\vphi_1,\ld,\vphi_n\ra = \pm z_i^{2^{n}}
|\pm \vec{z}\,^{2^n};\vphi_1,\ld,\vphi_n\ra.
\ee
{}$|\vec{z}\,^{2^n};\vphi_1,\ld,\vphi_n\ra$ are
superpositions of $2^n$ states $|\vec{z}\ra$ with $z_i$ on circles of
radius $|z_i|$. If all $\vphi_k$ are integer multiple of $\pi/2$ then
one gets the states $|\pm\vec{z}\ra$.
Independent parameters are $z,\vphi_1,\ld,\vphi_n$, therefore one could
also use the notation $|\vec{z};\vec{\vphi}^{(n)}\ra$ (as in section 3).
\\[0mm]

{\large {\it Theorem.}} If in equation (\ref{|vec{z}>1resolu})
$d\mu(\vec{z},\vec{z}^*) = F(|z_1|,\ld,|z_N|)\,d^2\vec{z}$ then

\be\lb{T}   
1 = \int
d\mu(\vec{z},\vec{z}\,^*)|\vec{z}\,^{2^n};\vphi_1,\ld,\vphi_n\ra\la\vphi_1,
\ld,\vphi_n;\vec{z}\,^{2^n}|,\quad n=0,1,2,\ld.
\ee

{\it Proof}. The theorem is valid for $n=0$ by construction. It is not
difficult to check directly, that it is valid for several $n>0$. Suppose
now that it is valid for $n-1$. Then we shall prove that it is valid for
$n$ as well. Indeed, using the definition (\ref{|z;vecvphi>}) and noting
that $-z_j^{2^n} = iz_j^{2^{n-1}}$ we obtain for the projectors in
(\ref{T}) the expression,

\bear\lb{projectors}  
|\vec{z}\,^{2^n};\vphi_1,\ld,\vphi_n\ra
\la\vphi_1,\ld,\vphi_n;\vec{z}\,^{2^n}| =
\cos^2\vphi_n
|\vec{z}\,^{2^{n-1}};\vphi_1,\ld,\vphi_{n-1}\ra
\la\vphi_1\ld,\vphi_{n-1};\vec{z}\,^{2^{n-1}}| \nn \\
+ \sin^2\vphi_n
|-\vec{z}\,^{2^{n-1}};\vphi_1,\ld,\vphi_{n-1}\ra
\la\vphi_1\ld,\vphi_{n-1};-\vec{z}\,^{2^{n-1}}| + \nn \\
i\cos\phi_n\sin\vphi_n |\vec{z}\,^{2^{n-1}};\vphi_1,\ld,\vphi_{n-1}\ra
\la\vphi_1,\ld,\vphi_{n-1};-\vec{z}\,^{2^{n-1}}| - \nn \\
i\cos\phi_n\sin\vphi_n |-\vec{z}\,^{2^{n-1}};\vphi_1,\ld,\vphi_{n-1}\ra
\la\vphi_1,\ld,\vphi_{n-1};\vec{z}\,^{2^{n-1}}|.
\eear
We substitute this expression into equation (\ref{|vec{z}>1resolu}) and then in the
second and in the last integral change the integration variables $z_i$ to
$z_j\exp[i\pi/2^{n-1}]$ (rotation on angle $\pi/2^{n-1}$). Then we note
that  under such rotation the eigenvalues $z_j^{2^{n-1}}$ of $A^{2^{n-1}}$
change the sign, i.e. $|\vec{z}\,^{2^{n-1}};\vphi_1,\ld,\vphi_{n-1}\ra \rar
|-\vec{z}\,^{2^{n-1}};\vphi_1,\ld,\vphi_{n-1}\ra$. This yields the
cancelation of the last two integrals and the coincidence of the first two
ones in view of the rotational invariance of the resolution unity measure
$d\mu(\vec{z},\vec{z}^*) = F(|z_1|,\ld,|z_N|)d^2\vec{z}$. We obtain

\bear  
\int
d\mu(\vec{z},\vec{z}\,^*)|\vec{z}\,^{2^n};\vphi_1,\ld,\vphi_n\ra\la\vphi_1,\ld,
\vphi_n;\vec{z}\,^{2^n}| =  \nn \\
\int
d\mu(\vec{z},\vec{z}\,^*)|\vec{z}\,^{2^{n-1}};\vphi_1,\ld,\vphi_{n-1}\ra
\la\vphi_1,\ld, \vphi_{n-1};\vec{z}\,^{2^{n-1}}|  = 1.
\eear
End the proof.

\subsection {A.3. On the uniqueness of the resolution unity measures
$d\mu(\vec{z},l,p,q)$ for $u(p,q)$ CS}

The resolution unity measure for a given continuous family of states is
generally not unique.  It could be unique if a certain constrain is
imposed on the class of admissible measures. For example, the requirement
of {\it invariance} of the measure on the group manifold under the group
action determines it uniquely \ci{BaRo}. As a result the resolution unity
measure for the group related CS is unique if it is invariant under the
group action.  For canonical SS families of noninvariant
resolution unity measures have been constructed in ref.  \ci{Trif93}.
Canonical SS minimize Schr\"odinger uncertainty relation and can be
regarded as group related CS for the semidirect product of $SU(1,1)$ and
Heisenberg--Weyl group \ci{Trif93}.

In this appendix we establish that the resolution unity measure for the
$u(p,q)$ CS (\ref{z-u(p,q)bgcs}) is uniquely determined by the requirement
the weight function $F(z_1,\ld,z_{N-1})$ to be a smooth function of
$|z_i|$ and independent of arg$z_i$, $i = 1,2,\ld,N-1$. Such is our weight
function in equation (\ref{upqresolu2}).

Suppose that there exists another function
$F^\pr(|z_1|,\ld,|z_{N-1}|;l,p,q)$ such that the new measure $d\mu^\pr =
F^\pr d^2\vec{z}$ resolves the unity $1_l$ as in equation  (\ref{upqresolu2}).
Then we should have

\be\lb{83}     
0 = \int d^2\vec{z}\l
[F(|\vec{\tld{z}_p}|,|\vec{\tld{z}_q}|;l,p,q)-
F^\pr(|z_1|,\ld,|z_{N-1}|;l,p,q)\r]
|\!|\vec{z};l,p,q\ra\la q,p,l;\vec{z}|\!|.
\ee
Substituting the expansion (\ref{z-u(p,q)bgcs}) of $|\!|\vec{z};l,p,q\ra$ and
integrating with respect to angles $\varphi_i=\arg z_i$ we obtain that the
difference function
$$\Phi(r_1,r_2,\ld,r_{N-1})\equiv
F(\tld{r}_p,\tld{r}_q;l,p,q)- F^\pr(|z_1|,\ld,|z_{N-1}|;l,p,q),$$
where
$\tld{r}_p\equiv |\vec{\tld{z}_p}|=\sqrt{r_1^2+...+r_p^2}$ and
$\tld{r}_q\equiv|\vec{\tld{z}_q}|=\sqrt{r^2_{p+1}+...+r^2_{N-1}}$, should
be orthogonal to the monomials

$$r_1^{2n_{1}+1}\ld r_{N-1}^{2n_{N-1}+1},\,\,\,
r_i=|z_i|,\,\, i=1,\ld,N-1,\,\, n_i=1,2,\ld.$$
Changing the integration variables and redenoting $r_i^2$ again as $r_i$
one can write this orthogonality in the form

\be\lb{ortogo2}   
\int_{0}^{\infty} dr_1\ld dr_{N-1}\Phi(r_1,\ld,r_{N-1}  )
r_1^{n_1}\ld r_{N-1}^{n_{N-1}} = 0,
\ee
where $ n_i = 1,2,\ld,\quad i=1,\ld,N-1$.  Equation (\ref{ortogo2})
implies that $\Phi(r_1,\ld,r_{N-1})$ is decreasing exponentially as the
total radius $r_1^2 +\ld + r_{N-1}^2$ tends to $\infty$. This means that
the integral $\int_0^\infty\Phi^2dr_1\ld dr_{N-1}$ is finite.  It follows
also from equation  (\ref{ortogo2}) that $\Phi$ is orthogonal to any function
$f(r_1\ld r_{N-1})$ which admits power expansion in terms of $r_i$,
$i=1,2,\ld N-1$,

\be\lb{ortogo3}     
\int_{0}^{\infty} dr_1\ld
dr_{N-1}\Phi(r_1,\ld,r_{N-1}) f(r_1\ld r_{N-1}) = 0.
\ee

This implies that $\Phi \equiv F-F^\pr = 0$ almost everywhere. Indeed, if
$\Phi\neq 0$ it must be nonpositive definite (in order to obey
(\ref{ortogo2})) and if $\Phi$ is well behaved (it is sufficient to be
continuous) we could find $f$ which is negative in the domains where
$\Phi<0$.  But then  we could not maintain (\ref{ortogo3}), unless
$F =F^\pr$ almost everywhere.  We suppose in (\ref{ortogo3}) that
the integral of the power series of $f$ is a sum of terms of the type of
(\ref{ortogo2}).  This is ensured if $\Phi$ is a smooth function
(i.e., all derrivatives finite) of $r_1,\ld, r_{N-1}$ (Our
$F$, equation  (\ref{F}), is such a function). In this case we can take in
(\ref{ortogo3}) $f=\Phi$.  Then we obtain that $F$ and $F^\pr$ should
coincide pointwise.  Thus the resolution unity measure (\ref{F}) is unique
within the set of smooth functions of $|z_1|,\ld,|z_{N-1}|$.

\subsection{A.4. Proof of the representation (\ref{K_nu}) of Bessel
function $K_{\nu}(z)$ }

In the case of $q=1$ ($p=N-1$) and $-l\geq 0$ our measure function $F$,
equation  (\ref{F}),  depends on $r_1,\ld,r_p$ through $|\vec{z}| =
[|z_1|^2+\ld+|z_{p}|^2]^{1/2} \equiv \tld{r}_p$ and it is a smooth and
positive function of $r_1,\ld,r_p$.  The measure function of ref.
\ci{Fujii} $F^\pr \sim \tld{r}_p^{-l-p+1}K_{-l-p+1}(2\tld{r}_p)$ is also
smooth and positive \ci{Stegun}. Therefore the difference $\Phi$ of these
two functions is smooth and in view of (\ref{ortogo2}) and the result of
the preceeding subsection they have to coincide pointwise.  This proves
formula (\ref{K_nu}) for ${\rm Im}z =0,\,\,{\rm Re}z > 0$ and $\nu=-l-p+1
= 0,\pm1,\ld$.

Let us consider the right hand side of (\ref{K_nu}) as a definition of a
new function $F(z;\nu)$, $z$ complex, $\nu$ real. The integral is
convergent for Re$z > 0$ and the function $F(z;\nu)$ is evidently analytic
with respect to $z$ and $\nu$.  The Bessel function $K_\nu(z)$ is analytic
and regular everywhere except of the negative half of the real line in
$z$-plain \ci{Stegun}.  We proved in the above that the two analytic
functions $F(z;\nu$ and $K_\nu(2z)$ ($\nu = 0,\pm1,\ld$) coincide on the
positive part of the real line of $z$.  Then they coincide in the whole
domain of analyticity in $z$-plain.  Numerical computations show that
formula (\ref{K_nu}) holds for complex $\nu$ as well. In conclusion let us
note that the integral in the right hand side of equation  (\ref{F})  correctly
defines (under replacements $|\vec{\tld{z}_p}|\rar z_1$,
$|\vec{\tld{z}_q}|\rar z_2$) analytic functions $F(z_1,z_2;l,p,q)$ of two
variables $z_1$ and $z_2$, Re$z_{1,2}>0$.  At $z_2=0$, $q=1$ we have
$F(z_1,0;l,p,1) = 2\pi^{-p}|z_1|^{1-l-p}K_{1-l-p}(2z_1)$.


\newpage

\centerline{ F I G U R E  \hs{5mm} C A P T I O N S}
\vs{8mm}

\begin{minipage}{14cm}
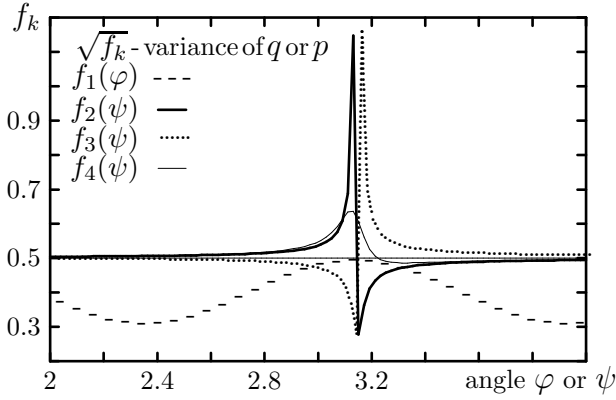
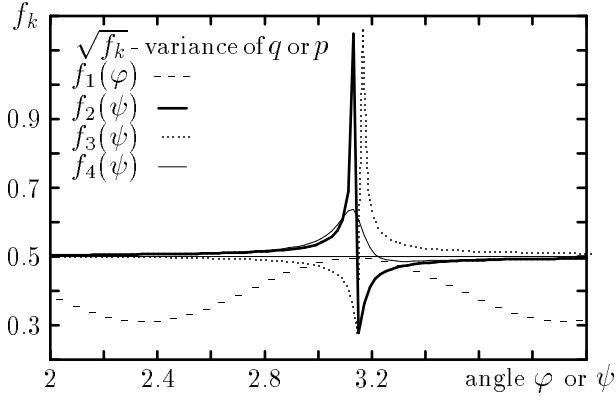
Fig. 1. Amplitude quadrature squeezing in
the $sp(N,R)$ BG CS $|\vec{\alf};\vphi\ra$ and the
superpositions
$|\vec{\alf},\phi,\psi\ra$, Eq.(61), for $r_i=|\alf_i|=0.05$.
$\Dlt\tld{q_i}=\Dlt\tld{q_i}(\tld{r},r_i,\tet_i,\phi,\psi)$,\,
$\tld{r} = |\vec{\alf}|$,\,\, $\tld{q}_i=q_i$ or $p_i$.
$f_1=\Dlt^2q_i(r_i,r_i,-\frac{\pi}{2},\vphi)$,\,
$f_2=\Dlt^2p_i(r_i,r_i,\frac{\pi}{4},0,\psi)$,
$f_3= \Dlt^2q_i(r_i,r_i,\frac{\pi}{4},0,\psi)$,\, $f_4 = \Dlt^2 p_i
(4r_i,r_i,\frac{\pi}{4},0,\psi)$.
$|\vec{\alf},\vphi\ra$ are weakly nonclassical states for every mode,\,\,
$|\vec{\alf},\phi,\psi\ra$ are strongly nonclassical.
\end{minipage}
\vs{5mm}

\begin{minipage}{14cm}
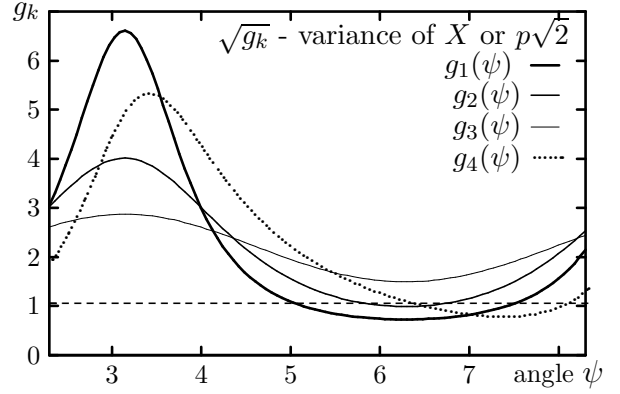
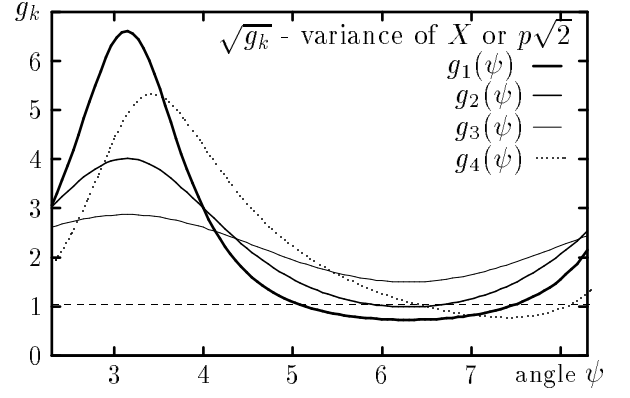
 Fig. 2. Squared amplitude quadrature squeezing in
 superpositions states $|\vec{\alf},\phi,\psi\ra$ for $r_i=0.8$.
 $\Dlt\tld{X_i}=\Dlt\tld{X_i}(\tld{r},r_i,\tet_i,\phi,\psi)$,\,
    $\tld{X}_i=X_i$ or $Y_i,\,\,\, \tld{r}=|\vec{\alf}|$.  $g_1 =
 \Dlt^2X_i(r_i,r_i,\frac{\pi}{4},0,\psi)=\Dlt^2Y_i(r_i,r_i,
  -\frac{\pi}{4},0,\psi)$,
 $g_2 = \Dlt^2X_i(1,r_i,\frac{\pi}{4},0,\psi)$,\,\,
$g_3=\Dlt^2X_i(1.2,r_i,\frac{\pi}{4},0,\psi)$,
 $g_4=2\Dlt^2p_i(r_i,r_i,\frac{\pi}{4},0,\psi)=2\Dlt^2q_i(r_i,r_i,
 -\frac{\pi}{4},0,\psi)$.
Joint $X$ and $p$ (or $Y$ and $q$) squeezing occurs in
 the interval $ -0.72 <\psi< -0.1 $.
\end{minipage} \vs{5mm}

\begin{minipage}{14cm}
 Fig. 3. Probabilities $p_n(\tld{r},\phi,\psi)$ to find $n$ photons
in the multimode superposition states $|\vec{\alf},\phi,\psi\ra$ as
functions of $\tld{r}=|\vec{\alf}|$ for different values of $\phi$ and
$\psi$.
$p_0=p_0(\tld{r},\frac{\pi}{2},\pi)$,\,
$p_1=p_1(\tld{r},\pi,-\frac{\pi}{2})$,\,
$p_2=p_2(\tld{r},0,\pi)$,\,
$p_3=p_3(\tld{r},\pi,\frac{\pi}{2})$,\,
$p_4=p_4(\tld{r},\frac{\pi}{4},\frac{\pi}{4})$,\,
$p_5=p_5(\tld{r},\pi,-\frac{\pi}{2})$.
\end{minipage} \vs{5mm}

\begin{minipage}{14cm}
\small Fig. 4. Oscillating photon number distributions
$p_n(\tld{r},\phi,\psi)$ in strongly nonclassical states
$|\vec{\alf},\phi,\psi\ra$ for different values of $\tld{r}$, $\phi$ and
$\psi$. $p_n1=p_n(0.8,0,7.3)$ ($Q\!>\!0$, $\Dlt p\!\geq\! 0.38$, $\Dlt
X\!\geq\! 0.73$),\, $p_n2=p_n(2.2,\pi,-\frac{\pi}{2})$ ($Q<0$),\,
$p_n3=p_n(0.55,5.0914,0)$ ($Q\!=\!-0$, $\Dlt p\!\geq\! 0.38$, $\Dlt
X\!>\!1$).
\end{minipage}  \vs{5mm}

\begin{minipage}{14cm}
Fig. 5. Nonoscillating photon number distributions in strongly
nonclassical states $|\vec{\alf},\phi,\psi\ra$  for different values of
$\tld{r}=|\vec{\alf}|$, $\phi$ and $\psi$.\,\, $p_n1 = p_n(0.55,2.246,0)$
\, ($Q<0$, $\Dlt q\geq 0.5$, $\Dlt X\geq 1$),\,\, $p_n2 =
p_n(0.55,2.33,0)$\, ($Q<0$,
$\Dlt q\geq 0.43$, $\Dlt X\geq 1$),\,\, $p_n3
= p_n(0.55,2.234384,0)$\, ($Q= +0$, $\Dlt q\geq 0.5$, $\Dlt X\geq 1$),\,\,
$p_n4 =$ Poisson distribution with $\la a^\dagger a\ra = 0.685$ as in $p_n3$.
\end{minipage}


\newpage

\begin{thebibliography}{99}

\bibitem{Trif94} Trifonov D.A. 1994 J. Math. Phys. {\bf 35}  2297;
        D.A. Trifonov, {\it Generalized intelligent states and $SU(1,1)$
        and $SU(2)$ squeezing}, Preprint INRNE-TH-93/4 (May 1993).

\bibitem{Vourd1} Brif C., Vourdas A. and Mann A. 1996 J.  Phys. A {\bf 29}
       5873 [E-print quant-ph/9607022].

\bibitem{Vourd2} Vourdas A., Brif C. and Mann A. 1996 J. Phys. A {\bf 29}
       5887 [E-print quant-ph/9606023].

\bibitem{Trif96a} Trifonov D.A. 1996, {\it Algebraic coherent states and
        squeezing}, E-print quant-ph/9609001.

\bibitem{Brif97a} Brif C. 1997 Int. J. Theor. Phys. {\bf 36} 1677
       [E-print quant-ph/9701003].

\bibitem{Trif97a} Trifonov D.A. J. Phys. A {\bf 30} (1997) 5941 [E-print
       quant-ph/9701018].

\bibitem{Fujii} Fujii K. and Funahashi K. 1997 J. Math. Phys. {\bf38} 4422
        [quant-th/9704011, E-print quant-th/9708041].

\bibitem{BG} Barut A.O. and Girardello L. 1971 Commun. Math. Phys. {\bf 21}
       41.

\bibitem{Brif96} Brif C. 1996 Ann. of Phys. {\bf 251} 180 [E-print
         quant-ph/9605006] .

\bibitem{KlSk} Klauder J.R. and  Skagerstam B.S. {\it Coherent States}
        (W.  Scientific, Singapore, 1985).

\bibitem{BaRo} Barut A.O. and Raszcka R. {\it Theory of Group
        Representations and Applications} (Polish Publishers, Warszawa,
        1977).

\bibitem{Mosh} Castanos O., Chacon B., Moshinsky M. and Quesne C.
    1985 J. Math. Phys. {\bf 26} 2107.

\bibitem{Tod} Todorov I.T., in {\it High Energy Physics and
       Elementary Particle Theory} ("Naukova Dumka", Kiev, 1967) (In
       Russian).

\bibitem{cats1} Dodonov V.V., Malkin I.A. and Man'ko V.A. 1974 Physica
          {\bf 72} 597;
       Yurke B. and Stoler D.A. 1986 Phys. Rev. Lett. {\bf 57} 13;
       Janszky J. and Vinogradov A.K. 1990 Phys. Rev. Lett. {\bf 64}
       277;
       Johsi J. and Singh M. 1995 J. Mod. Opt. {\bf 42}(4), 775;
       Moya-Cessa H. and Vidiella-Baranco A. 1995 J.  Mod. Opt. {\bf
       42}(7), 1547;
       Haroche S. 1995 N. Cimento B {\bf 110} 545;
       Goetsch P., Graham R. and Haake F. 1995 Phys. Rev. A{\bf 51} 136.

\bibitem{cats2} Ansari N.A. and Man'ko V.I. 1994 Phys. Rev. A{\bf 50} 1942;
       Dodonov V.V., Man'ko V.I. and Nikonov D.E. 1995 Phys. Rev. A {\bf 51}
       3328 (1995).

\bibitem{Agar} Agarwal G.S. 1988 J. Opt. Soc. Am. B {\bf 5} 1940.

\bibitem{Hill} Bergou J.A., Hillery M. and Yu D. 1991 {\it Phys. Rev.}
      A {\bf 43} 515.

\bibitem{Arv1} Arvind, Mukunda N. and Simon R. 1997 Phys. Rev. A {\bf56}
    5042 [E-print quant-ph/9611014];
    Arvind, Mukunda N. and Simon R. 1998 J. Phys. A {\bf31} 565.

\bibitem{Klysh} Klyshko D.N. 1996 Phys. Lett. A {\bf213} 7; Klyshko D.N.
    1996 Usp. Fiz. Nauk. {\bf166} 613 [Sov. Phys. Usp. {\bf39} (1966) 573].

\bibitem{Ger2} Gerry C.C. and Grobe R. 1995 Phys. Rev. A {\bf 51} 1698.

\bibitem{Prak} Prakash G.S. and  G.S. Agarwal  1995 Phys. Rev. A {\bf 52}
		2335.

\bibitem{Ger1} Gerry C.C. and Grobe R. 1995 Phys. Rev. A {\bf 51} 4123.

\bibitem{Shanta} Shanta P., Chaturvedi S. and Srinivasan V. 1996 Mod.
    Phys. Lett. A {\bf11} 2388 [E-print quant-ph/9608034].

\bibitem{Arv2} Simon R., Selvadoray M., Arvind and Mukunda N.
        {\em Necessary and Sufficient Classicality Conditions on
		Photon Number Distributions}, E-print quant-ph/9709030.

\bibitem{NT} Nieto N.N. and Truax D.R. 1993  Phys. Rev. Lett. {\bf 71} 2843.

\bibitem{Stegun} {\it Handbook of Mathematical  Functions}, edited  by M.
    Abramowitz and I.A. Stegun ({\it National bureau of standards},  1964)
    (Russian translation, "Nauka", M. 1979).

\bibitem{Basu} Basu D. 1992 J. Math. Phys. {\bf 33} 114.

\bibitem{Ben} Brif C. and Ben-Aryeh Y. 1994 J. Phys. A {\bf 27} 8185.

\bibitem{Trif97b} Trifonov D.A. 1997, {\it On the squeezed states for $n$
    observables}, E-print quant-ph/9705001 (to appear in Phys. Scripta).

\bibitem{Trif95} Trifonov D.A. 1995, {\it Uncertainty matrix, multimode
	squeezed states and generalized even and odd coherent states},
	Preprint INRNE-TH-95/5 (Abstract in {\it 1995 INRNE Annual Report},
	p. 57 (Sofia, 1996)).

\bibitem{Trif96b} Trifonov D.A. 1996, {\it Schr\"odinger intelligent
    states and linear and quadratic amplitude squeezing},
     E-print quant-ph/9609017.

\bibitem{Puri} Puri R.R. and Agarwal G.S. 1996 Phys. Rev. A {\bf 53} 1786.

\bibitem{Perina} Luis A. and Perina J. 1996 Phys. Rev. A {\bf 53} 1886.

\bibitem{Ger3} Hach III E.E. and Gerry C.C. 1992 J. Mod. Opt. {\bf 39}
       2501;
      Buzek V., Jeh I. and Quang T. 1990 J. Mod. Opt. {\bf 37} 159.

\bibitem{Vacc} Vaccaro J.A. and Orlowski A. 1995 Phys. Rev. A {\bf 51} 4172.

\bibitem{Bose} Bose S., Jacob K. and Knight P.L. 1997 Phys. Rev. A {\bf56}
       4175-86 [E-print quant-ph/9708002].

\bibitem{Jans} Janszky J., Domokos P, Szabo S. and Adam P. 1995
        Phys. Rev. A {\bf 51}(5), 4191;
        Szabo S., Adam P., Janszky J. and Domokos P. 1996 Phys. Rev.
        A {\bf 53}(4).

\bibitem{Rob} H.R. Robertson,  Phys. Rev. {\bf 46} 794 (1934).

\bibitem{Titulaer} Titulaer U. and Glauber R. 1965 Phys. Rev. {\bf 145}
      1041.

\bibitem{Mandel} Mandel L. 1983 Phys. Rev. Lett. {\bf 49} 136;
        Mandel L. 1979 Opt. Lett., {\bf 4} 205.

\bibitem{Walls} Walls D.F. 1983 Nature (London), {\bf 306} 141;
      Loudon R. and Knight P. 1987 J. Mod. Opt. {\bf 34} 709.
    {\sm [Eigenstates of complex combinations $u_{ki}a_i + v_{ki}a^\dg_i$,
     i.e. multimode canonical SS or canonical Robertson IS\ci{Trif97a},
   have been considered in \ci{XMa}, also in Malkin I.A.,
   Man'ko V.I.  and Trifonov D.A., Phys. Lett. A {\bf 30} (1969) 413; N.
   Cimento A {\bf 4} (1971) 773; Holz A., Lett. N. Cimento A {\bf 4}
   (1970) 1319]}.

\bibitem{XMa} Ma X and Rhodes W 1990, Phys. Rev. {\bf 41} 4624.

\bibitem{Brif97b} Brif C. and Mann A. 1997 Quant. Semiclass. Opt. {\bf9}
    899 [E-print quant-ph/9707009].

\bibitem{Marian} Marian P. 1997 Phys. Rev. A {\bf 55} 3051.

\bibitem{Nagel} Nagel B. 1997, {\it Higher power SS, Jacobi	matrices,
		and the hamburger moment problem}, E-print quant-ph/9711028.

\bibitem{Trif93} Trifonov D.A. 1993 J. Math. Phys. {\bf 34}  100.

\end{thebibliography}
\end{document}